\documentclass[12pt]{article}
\usepackage{a4,amssymb,verbatim,amsmath}

\newtheorem{theorem}{Theorem}[section]

\newtheorem{remark}[theorem]{Remark}
\newtheorem{corollary}[theorem]{Corollary}
\newtheorem{proposition}[theorem]{Proposition}

\newtheorem{ex}{Example}[section]
\newenvironment{example}{\begin{ex}\rm}{ \hfill $\Diamond$ \end{ex}
        \vskip4pt}
\newtheorem{ass}{Assumption}[section]

\numberwithin{equation}{section}

\begin{document}

\def\sso#1{\ensuremath{\mathfrak{#1}}}  
\def\ssobig#1{{{#1}}}

\newcommand{\ddx}{\partial \over \partial x}
\newcommand{\ddy}{\partial \over \partial y}

\begin{center}
{\bf \Large Lie group classification of first-order delay ordinary differential equations}
\end{center}

\bigskip

\begin{center}
{\large Vladimir A. Dorodnitsyn}$^{*}$,
{\large Roman Kozlov}$^{**}$, \\
{\large  Sergey V. Meleshko}$^{\dag}$
{\large and Pavel Winternitz}$^{\ddag}$

\bigskip

${}^{*}$
Keldysh Institute of Applied Mathematics, Russian Academy of Science, \\
Miusskaya Pl.~4, Moscow, 125047, Russia; \\
e-mail address: DorodnitsynVA@gmail.com \\

$^{**}$ Department of Business and Management Science, \\ Norwegian
School of Economics, Helleveien 30, 5045, Bergen, Norway; \\
 {e-mail: Roman.Kozlov@nhh.no}\\

${}^{\dag}$
School of Mathematics, Institute of Science, \\ Suranaree University of Technology, 30000, Thailand; \\
e-mail address: sergey@math.sut.ac.th \\

$^{\ddag}$ Centre de Recherches Math\'ematiques
and
D\'epartement de math\'ematiques et de statistique,
Universit\'e de Montr\'eal,
Montr\'eal, QC, H3C 3J7, Canada; \\
e-mail address: wintern@crm.umontreal.ca \\

\end{center}

\bigskip

\begin{center}
\end{center}

\bigskip

\begin{center}
{\bf Abstract}
\end{center}
\begin{quotation}

 A group classification of first-order delay ordinary differential
equation (DODE) accompanied by an equation for delay parameter
(delay relation) is presented. A subset of such systems 
(delay ordinary differential systems or DODSs)
which consists of linear DODEs and solution independent delay
relations have infinite-dimensional symmetry algebras, as do
nonlinear ones that are linearizable by an invertible transformation
of variables.
Genuinely nonlinear DODSs have symmetry algebras of
dimension $n$, $0 \leq n \leq 3$. It is shown how exact analytical
solutions of invariant DODSs can be obtained using symmetry
reduction.
\end{quotation}

\bigskip

\eject

\bigskip
\section{\large \bf  Introduction}     \label{ Introduction}

Lie groups have provided efficient tools for studying ordinary and
partial differential equations since their introduction in the
seminal work of Sophus Lie in the late 19th century~\cite{Lie15,
Lie72SW, bk:Lie[1891b], Lie14}. The symmetry group of a differential
equation transforms solutions into solutions while leaving the set
of all solutions invariant. The symmetry group can be used to obtain
new solutions from known ones and to classify equations into
equivalence classes according to their symmetry groups. It can also
be used to obtain exact analytic solutions that are invariant under
some subgroup of the symmetry group, so called "group invariant
solutions". Most solutions of the nonlinear differential equations
occurring in physics and in other applications were obtained in this
manner. Applications of Lie group theory to differential equations,
known as group analysis, is the subject of many books and review
articles~\cite{Ovsiannikov1982,Olver1986,Ibragimov1985,Bluman1989,
bk:HandbookLie, Gaeta1994}.

More recently applications of Lie group analysis have been extended
to discrete equations~\cite{Maeda1, Maeda2,Dor_1, Dor_2, Dor_3,
DorKozWin, Levi, Levi2006,  DKW2000,  Quisp,  Wint, Dorodnitsyn2011,
Vinet, Hydon}, both difference and differential-difference ones. The
applications are the same as in the case of differential equations.
One can classify discrete equations into symmetry classes, obtain
invariant solutions and do everything that one does for differential
equations.

The present article is part of a research program the aim of which
is to extend the application of group analysis to another type of
equations, namely differential delay equations.  These are equations
involving a field $y(x)$ and its derivatives, all evaluated not only
at the point  $x$, but also at some preceding points $x_-$. In
general, the independent and dependent variables $x$ and $y$ can be
vectors. In this article we restrict to the case when both are
scalars.  Thus we have just one equation and one independent
variable, the case of a delay ordinary differential equation (DODE).

Differential delay equations play an important role in the
description of many phenomena in physics, chemistry, engineering,
biology, economics, etc.. They occur whenever the state of  a system
depends on previous states. For instance a patients blood pressure
$y(t)$ may depend on the blood pressure $y(t_-)$ at the moment $t_-$
when medication had been delivered. The state of a bank account may
depend on transactions made at previous times, etc.

A sizable literature exists on delay differential equations and
their applications \cite{ Elgolts1964,  Hale1971, Kolmanovskii,
bk:Myshkis1972,Driver1977, Myshkis1989, bk:Richard[2003], bk:Wu1996,
Erneux}. In particular for applications in many fields and for a
useful list of references we turn to the book \cite{Erneux} by
T.~Erneux. Studies have been devoted to the existence and stability
of solutions, to the general properties of solutions and to
questions of periodicity and oscillations.

Nearly all known solutions to delay differential equations are
numerical ones. Analytical solutions have been obtained by
postulating a specific parameterized form and then adjusting the
parameters to satisfy the equation~\cite{bk:PolyaninZhurov[2014]a,
bk:PolyaninZhurov[2014]b, bk:PolyaninZhurov[2014]aa}. For a specific
application in biology, see~\cite{Application0, Craig,    
Application1, Application2}. For earlier work on
symmetry analysis for delay-differential equations
see~\cite{bk:PrapartMeleshko[2008]}.

In this article we restrict to first order delay ordinary
differential equations (DODEs) of a specific type, classify them into
symmetry classes and present some invariant solutions.

The article is organized as follows.
In Section~\ref{Formulation_of_the_problem} we formulate the
problem. We define a DODS, that is a DODE together with a delay
equation that specifies the position of the delay point $x_-$. In
Section~\ref{generaltheory} we provide the general theory
and outline the method for classifying DODSs.
The Lie group classification of first-order DODSs
is given in Section~\ref{Classification}.
Section~\ref{linearsection}   is devoted to linear DODEs with
solution independent delay relations. They all admit
infinite-dimensional symmetry groups.
In Section~\ref{invariantsolutions1}    we show how
symmetries can be used to find particular (namely group invariant)
solutions of DODEs.
Finally, the conclusions are presented in
Section~\ref{Conclusions}.
The results of the classification of nonlinear  first-order DODSs of
the considered form are summed up in Table 2.

\section{\large \bf  Formulation of the problem}

\label{Formulation_of_the_problem}

The specific purpose of this article is to perform a symmetry classification
of first-order delay
ordinary differential equations (DODEs)
\begin{equation}  \label{DODE}
\dot{y} = f ( x, y, y_- ),
\qquad
{ \partial f  \over  \partial y_- } {\not\equiv}  0  ,
\qquad
x \in I ,
\end{equation}
where $I \subset  \mathbb{R} $ is some finite or semi{finite} interval.
We will be interested in symmetry properties of this DODE,
which is considered locally, independently of initial conditions.
For equation~(\ref{DODE})  we have to specify the delayed point $x_-$
where the delayed function value $ y_- = y(x_-) $ is taken,
otherwise the problem is not fully determined.
Therefore, we supplement the DODE with a delay relation
\begin{equation}    \label{delay}
x_- = g ( x, y, y_- ) ,
\qquad
x_- <  x ,
\qquad
g  ( x, y, y_- ) {\not  \equiv}   \mbox{const} .
\end{equation}

The two equations~(\ref{DODE}) and~(\ref{delay}) together
will be called a {\it delay ordinary differential system} (DODS).
Here $f$ and $g$ are arbitrary smooth functions.
Sometimes for convenience we shall write~(\ref{delay})
in the equivalent form
$$
\Delta x  = x - x_- = \tilde{g} ( x, y, y_- ) ,
\qquad
\tilde{g} ( x, y, y_- ) =   x -  g ( x, y, y_- ) .
$$
In most of the existing literature  the delay
parameter $ \Delta x $ is considered to be constant
\begin{equation}    \label{consntantdelay}
 \Delta x  = \tau  > 0 ,
 \qquad
 \tau  = \mbox{const} .
\end{equation}
An alternative is to impose a specific form of the function~(\ref{delay})
to include some physical features of the delay $\Delta x$.

In the classification that we are performing we will find all
special cases of functions $f$ and $g$, when the DODS under
consideration will possess a nontrivial symmetry group. For all such
cases of $f$ and $g$ the corresponding group will be presented.


We will be interested in group transformations, leaving
the Eqs.~(\ref{DODE}),(\ref{delay}) invariant.
That means that the transformation will transform solutions of the DODS into solutions.
They leave the set of all solutions invariant.
Let us stress that we need to consider these two equations together.
This makes our approach similar to one of the approaches
for considering symmetries of discrete equations
(see, for example,~\cite{Dorodnitsyn2011, DKW2000}),
where the invariance is required for both  the discrete equation and the equation for the lattice
on which the discrete equation is considered.
Here, we have the DODE~(\ref{DODE})  instead of a discrete equation
and the delay relation~(\ref{delay})  instead of a lattice equation.

In order to solve a DODE on some interval $I$
one must add initial conditions to the DODS~(\ref{DODE}),(\ref{delay}).
Contrary to the case of ordinary differential equations,
the initial condition must be given by a function $ \varphi (x)$
on an initial interval $ I_0 \subset  \mathbb{R} $, e.g.
\begin{equation}    \label{initialvalues}
 y (x)   = \varphi (x), \qquad
x \in [ x_{-1} , x_0 ]  .
\end{equation}
For $ x - x_-  = \tau  = x_0 - x_{-1} $ constant this leads to the
{\it method of steps}~\cite{Myshkis1989} for solving the DODE
either analytically or numerically.
Thus for $ x_0 \leq x \leq x_1 = x_0 + \tau $ we replace
$$
y _- = y ( x _- ) =   \varphi (x - \tau )
$$
and this reduces the DODE~(\ref{DODE}) to an ODE
\begin{equation}
\dot{y} (x)  = f ( x, y (x) , \varphi (x - \tau ) ),
\qquad
x_0 \leq x \leq x_1 = x_0  + \tau
\end{equation}
which is solved with initial condition $ y(x_0) = \varphi (x _0  )$.

On the second step we consider the same procedure:
we solve the ODE
\begin{equation}
\dot{y} (x)  = f ( x, y (x) , y (x - \tau ) ),
\qquad
x_1 \leq x \leq x_2 = x_1 + \tau ,
\end{equation}
where $y (x - \tau ) $ and the initial condition are known from the first step.
Thus we continue until we cover the entire interval $I$,
at each step solving the ODE with input from the previous step.
This procedure provides a solution that is in general continuous in the points $ x_n = x_0 + n \tau $
but not smooth.

The condition~(\ref{consntantdelay})
is very restrictive and rules out most of the symmetries that could be present for ODEs.
Hence we consider the more general case of the DODS~(\ref{DODE}),(\ref{delay})
and we must adapt the method of steps to this case.
We again use the initial condition~(\ref{initialvalues}),
which is required to satisfy the condition
$$
 x_{-1} = g (  x_0 ,   \varphi ( x_0 ),    \varphi ( x_{-1} )  )  .
$$
We will also assume that
\begin{equation}       \label{conditions}
x_{-1} \leq   x_-   < x  .
\end{equation}


On the first step we solve the system
\begin{equation}
\dot{y} (x) = f  ( x,  y (x) , \varphi ( x_- )   ) ,
\end{equation}
\begin{equation}
 x_-  = g ( x , y (x) , \varphi ( x_- )  )  ,
\end{equation}
for $y(x)$  and $x_- (x)$ from the point $x_0$
with initial condition  $ y(x_0) = \varphi ( x_0 ) $.
The system is solved forward till a point $x_1$ such that $ x_0 = g (  x_1 ,   y  ( x_1 ),    y ( x_{0} )) $.
Thus, we obtain  the solution $y(x)$ on the interval  $[ x_{0} , x_1 ] $.
In the general case of delay relation~(\ref{delay})
point $x_1 $ can depend on the particular solution $y(x)$ generated by the initial values~(\ref{initialvalues}).

Once we know the solution on the interval   $[ x_{0} , x_1 ] $,
we can proceed to the next interval $[ x_{1} , x_2 ]$
such that $ x_1 = g (  x_2 ,   y  ( x_2 ),    y ( x_{1} )  )$ and so on.
Thus the method of steps for the solution of the initial value problem introduces
a natural sequence of intervals
\begin{equation}   \label{intervals}
[ x_{n} , x_{n+1} ] ,      \qquad      n = -1, 0, 1, 2, ...
\end{equation}
We emphasize that the introduction of the intervals~(\ref{intervals}) is not a discretization.
The variable $x$ varies
continuously over the entire region where Eqs.~(\ref{DODE}),(\ref{delay})
are defined as does the dependent variable $y$.

\section{Construction of invariant DODS}

\label{generaltheory}

In this section we describe the method to be used to classify all
first-order delay ordinary differential systems~(\ref{DODE}),(\ref{delay})
that are invariant under some nontrivial Lie group of point transformations
into conjugacy classes under local diffeomorphisms.
For each class we propose a representative DODS.





We realize the Lie algebra $L$
of the point symmetry group $G$  of the system~(\ref{DODE}),(\ref{delay})
by vector fields of the same form as in the case of ordinary differential equations,
namely
\begin{equation}    \label{operator1}
X _{\alpha}  =   \xi  _{\alpha}   (x,y)  { \ddx}
+ \eta  _{\alpha}  (x,y)  { \ddy}  ,
\qquad
\alpha  = 1, ..., n .
\end{equation}
The prolongation of these vector fields acting on the system~(\ref{DODE}),(\ref{delay})
will have the form
\begin{equation}    \label{prolongation}
\mbox{\bf pr} X  _{\alpha}
=    \xi _{\alpha}  { \ddx}  + \eta _{\alpha}   { \ddy}
+   \xi  _{\alpha}   ^-   {\partial  \over \partial x_-}  + \eta _{\alpha}   ^-  {\partial   \over \partial y_-}
+   \zeta  _{\alpha}   {\partial  \over \partial \dot{y} }
\end{equation}
with
\begin{equation*}
\xi _{\alpha}   = \xi  _{\alpha} (x,y) ,
\qquad
\eta _{\alpha}  = \eta   _{\alpha}  (x ,y )  ,
\end{equation*}
\begin{equation*}
\xi _{\alpha} ^-    = \xi   _{\alpha}  (x_- ,y_-)  ,
\qquad
\eta  _{\alpha} ^-    = \eta   _{\alpha} (x_- ,y_-)  ,
\end{equation*}
\begin{equation*}
\zeta   _{\alpha}   (x,y, \dot{y} )  = D  ( \eta  _{\alpha}  )  - \dot{y}   D  ( \xi   _{\alpha} )  ,
\end{equation*}
where $D$ is the total derivative operator.
Let us note that  Eq.~(\ref{prolongation}) combines
prolongation for shifted discrete variables $\{ x_- , y_- \} $~\cite{Dorodnitsyn2011, DKW2000}
with standard prolongation for  the continuous derivative $\dot{y}$~\cite{Ovsiannikov1982, Olver1986}.

All finite-dimensional complex Lie algebras of vector fields of the
form~(\ref{operator1}) were classified by S.Lie~\cite{Lie15, Lie14}.
The real ones were classified more recently in~\cite{Gonzalez1992}
(see also~\cite{Nesterenko1, Nesterenko2}). The classification is
performed under the local group of diffeomorphisms
\begin{equation*}     
\bar{x} = \bar{x} ( x, y) ,
\qquad
\bar{y} = \bar{y} ( x, y) .
\end{equation*}

In the next sections we construct the invariant delay ordinary differential equations
supplemented by invariant delay relations,  proceeding by dimension of the Lie algebra.
In the paper we will  use  following notations
$$
\Delta x = x - x_- ,
\qquad
\Delta y = y - y_- .
$$
In the ``no delay'' limit we have $ \Delta x \rightarrow 0 $, $ \Delta
y \rightarrow 0 $.

We shall run through the list of representative algebras
given in in~\cite{Gonzalez1992}
and for each of them construct all group invariants in the jet space with local coordinates
$(x  , y , x_- , y_- , \dot{y}) $.
We shall construct both strong invariants (invariant  in the entire space)
and weak invariants (invariant on some manifolds).
The invariants will be used to write invariant differential delay systems~(\ref{DODE}),(\ref{delay}).
The usual restriction $x_-  = x - \tau $,  $\tau = \mbox{const}$
excludes most symmetries. We however use the delay relation~(\ref{delay}) instead.
The situation is now similar to that of ordinary difference schemes
considered in the earlier article~\cite{DKW2000}.

Let us assume that a Lie group $G$ is given
 and that its Lie algebra $L$
is realized by the vector fields of the form~(\ref{operator1}).
If we wish to construct a first-order DODE with a delay relation
that are  invariant under this group, we proceed as follows.
We choose a basis of the Lie algebra, namely $\{ X _{\alpha} , \alpha = 1, ..., n \}$,
and impose the equations
\begin{equation}
\mbox{\bf pr} X  _{\alpha}    \Phi ( x, y ,x_- , y_- ,\dot{y} ) = 0 ,
\qquad
 \alpha = 1, ... , n  ,
\end{equation}
 with  $\mbox{\bf pr} X  _{\alpha}  $ given by Eq.~(\ref{prolongation}).
Using the method of characteristics, we obtain a set of elementary invariants $ I_1, ... , I_k$.
Their number is
\begin{equation}
k =  \mbox{dim} \  M - (   \mbox{dim} \   G  -    \mbox{dim} \   G  _0 )   ,
\end{equation}
where $M$ is the manifold that $G$  acts on and $G _0 $  is the
stabilizer of a generic point on $M$. In our case we have $  M  \sim
( x, y ,x_- , y_- ,\dot{y} )  $ and hence $ \mbox{dim} \   M  = 5 $.

Practically, it is convenient to express the number of invariants as
\begin{equation}
k =   \mbox{dim} \   M -  \mbox{rank} \  Z ,
\qquad
k   \geq  0   ,
\end{equation}
where $Z$ is the matrix
\begin{equation}
Z = \left(
\begin{array}{ccccc}
\xi_1    & \eta_1   &  \xi _1 ^-  &  \eta  _1 ^-   & \zeta _1   \\
\vdots  & & & &  \\
\xi_n    & \eta_n   &  \xi _n ^-  &  \eta  _n ^-   & \zeta _n   \\
\end{array}
\right)   .
\end{equation}
The rank of $Z$ is calculated at a generic point of $M$.

The invariant DODE and delay relation are written as
\begin{equation}   \label{implicit1}
F (   I_1, ... , I_k ) = 0   ,
\end{equation}
\begin{equation}   \label{implicit2}
G (   I_1, ... , I_k ) = 0   ,
\end{equation}
 where $F$ and $G$ satisfy
\begin{equation*}
\mbox{det}  \  {\partial ( F , G )  \over \partial ( \dot{y}  , x_- ) } \neq 0   .
\end{equation*}
Note that Eqs.~(\ref{implicit1}),(\ref{implicit2}) can be rewritten in the form~(\ref{DODE}),(\ref{delay}).
Equation~(\ref{implicit1}) with delay relation~(\ref{implicit2})
obtained in this manner are "strongly invariant",
i.e.   $ \mbox{\bf pr}   X  _{\alpha}  F = 0 $ and $\mbox{\bf pr}  X  _{\alpha}  G = 0 $
are satisfied identically.

Further invariant equations are obtained if the rank of $Z$
is less than maximal on some manifold described by the equations
\begin{equation}     \label{implicit3}
F (   x, y ,x_- , y_- , \dot{y}  )  = 0     ,
\end{equation}
\begin{equation}    \label{implicit4}
G (   x, y ,x_- , y_- ,  \dot{y} )  = 0   ,
\end{equation}
\begin{equation*}
\mbox{det}  \  {\partial ( F , G )  \over \partial ( \dot{y}  , x_- ) } \neq 0   ,
\end{equation*}
which satisfy the conditions
\begin{equation}     \label{invariance3}
\left.
\mbox{\bf pr} X  _{\alpha}  F
\right| _{F = 0, \ G = 0 }  = 0   ,
\qquad
\alpha = 1, ..., n  ,
\end{equation}
\begin{equation}     \label{invariance4}
\left.
\mbox{\bf pr} X  _{\alpha}  G
\right| _{F = 0, \ G = 0 }  = 0  ,
\qquad
 \alpha = 1, ..., n  .
\end{equation}
Thus we obtain "weakly invariant" DODS~(\ref{implicit3}),(\ref{implicit4}),
i.e. equations~(\ref{invariance3}) and~(\ref{invariance4})  are satisfied on the solutions of
the system $ F = 0 $, $G = 0$.

Note that we must discard trivial cases when the obtained equations~(\ref{DODE}),(\ref{delay})
do not satisfy the conditions
\begin{equation}    \label{DODS_conditions}
{ \partial f  \over  \partial y_- }  {\not\equiv}  0 ,
\qquad
x_- <  x
\qquad
\mbox{or}
\qquad
g ( x, y, y_- ) {\not\equiv}  \mbox{const}
\end{equation}
as expected for a DODS.


We shall make use of two different existing classifications
of real finite dimensional Lie algebras.
One is the complete classification of the finite-dimensional Lie sub{algebras}
of  $ \mbox{diff} ( 2, \mathbb{R}  ) $,
realized by vector fields of the form~(\ref{operator1}).
This classification into conjugacy classes under the group of inner automorphisms,
i.e. under arbitrary local transformations of variables was presented in~\cite{Gonzalez1992}.
An earlier classification of sub{algebras} of $ \mbox{diff} ( 2, \mathbb{C}  ) $
is due to Sophus Lie~\cite{Lie15, Lie14}.
The other is the classification of low-dimensional Lie algebras into isomorphism classes.
This classification is complete for  $  \mbox{dim} \ L  \leq 6$.
The work is due to many
authors~\cite{Lie72SW, Morozov1958, Mubarakzjanov1963, Mubarakzjanov1966, Patera1976, Turkowski1992}.
The classification together with many results on the classification of certain series
of solvable Lie algebras are summed up in the book~\cite{SWbook}.

These two classifications are combined together for algebras satisfying
 $   1 \leq    \mbox{dim} \ L    \leq 4 $ in Table~1.
The notations are explained in the table caption.


Indecomposable Lie algebras precede the decomposable ones
(like  $ 2 {\sso {n}}_{1,1} $ or   $ {\sso {n}}_{1,1} \oplus  {\sso {s}}_{2,1} $)
in the list.
Isomorphic Lie algebras can be realized in more than one manner by vector fields.
Already at dimension $ \mbox{dim} \ L    = 2 $ we see that     $  {\sso {s}}_{2,1} $ and  $ 2 {\sso {n}}_{1,1} $
have two inequivalent realizations each.
Thus the vector fields in $ \mbox{\bf A}_{2,1} $ and $ \mbox{\bf A}_{2,3} $ are
``linearly connected'' ($X_1$ and $X_2$ are both proportional to $ \partial / \partial y$).
In $ \mbox{\bf A}_{2,2} $ and $ \mbox{\bf A}_{2,4}  $  they span the entire tangent space
$ \{  \partial / \partial  x , \partial / \partial  y  \} $.
Such operators $X_1$ and $X_2$ are called ``linearly non{connected}''.
In dimension 3 for instance $ {\sso {sl}} ( 2, \mathbb{R}  ) $ is represented in 4 different inequivalent manners.
For nilpotent Lie algebras elements of the derived algebra precede a semicolon,
e.g. $ X_1$,   $ X_2$ in    $ {\sso {n}}_{4,1} $.
For solvable Lie algebras the nil{radical} precedes a semicolon,
e.g. $ X_1$,   $ X_2$,    $ X_3$ in    $ {\sso {s}}_{4,6} $.

\section{Classification of invariant DODEs}         \label{Classification}

In this section we construct invariant DODEs with invariant delay relations.
For this purpose we need to find invariants in the space
$ ( x, y, x_-, y_- , \dot{y}  ) $
as explained in the previous section.

\subsection{Dimension 1}

We start with the simplest case of a symmetry group, namely a one-dimensional one. Its Lie
algebra is generated by one vector field of the form~(\ref{operator1}).
By an appropriate change of variables
we take this vector field into its rectified form
(locally in a nonsingular point $(x,y)$).
Thus we have $ \mbox{\bf A}_{1,1} $:
\begin{equation}     \label{Dcase11}
X_1 = { \ddy}    .
\end{equation}

In order to write a first-order DODE and a delay relation invariant under this group
we need the invariants annihilated by the prolongation~(\ref{prolongation}) of $X_1$ to the prolonged
space $ ( x , y , x_-  , y_-  , \dot{y} )$.
A basis for the invariants is
$$
I_1 = x  ,
\qquad
I_2 = x_-  ,
\qquad
I_3  = \Delta y  ,
\qquad
I_4 = \dot{y} .
$$
The most general first-order DODE with delay relation invariant under the corresponding group
can be writen as
\begin{equation}    \label{DODEcase11}
\dot{y} = f \left( x, { \Delta y  \over \Delta x } \right)  ,
\qquad
 \Delta x  =   g ( x, \Delta y  )   ,
\end{equation}
where $f$ and $g$  are arbitrary functions.
Writing $\dot{y} = h  ( x, \Delta y )  $,  $\Delta x  =   g ( x, \Delta y  ) $
would obviously be equivalent.

\subsection{Dimension 2}


$ \mbox{\bf A}_{2,1} $:
The non-Abelian Lie algebra  ${\sso {s}}_{2,1}$   with linearly connected basis elements
\begin{equation}      \label{Dcase24}
X_1 = { \ddy}   ,
\qquad
X_2 =  y  { \ddy}
\end{equation}
provides us with the invariants
$$
I_1 = x   ,
\qquad
I_2 = x_- ,
\qquad
I_3 =  { \dot{y}   \over   \Delta  y }   .
$$
The most general invariant DODS can be written as
\begin{equation}       \label{DODEcase24}
\dot{y}   =    f (x)   {  \Delta y     \over    \Delta  x   } ,
\qquad
   x_-    =      g (x)    .
\end{equation}
Here and below we will assume that conditions~(\ref{DODS_conditions})
hold. For the equations~(\ref{DODEcase24}) they imply
\begin{equation*}
f(x) {\not\equiv} 0 , 
\qquad
   g (x) < x ,
\qquad
 g(x) {\not\equiv}  \mbox{const}    .
\end{equation*}



\bigskip


$\mbox{\bf A}_{2,2}$:
The non-Abelian Lie algebra ${\sso {s}}_{2,1}$ with non{connected} elements
\begin{equation}      \label{Dcase23}
X_1 = { \ddy}   ,
\qquad
X_2 = x { \ddx}  +   y  { \ddy}
\end{equation}
yields a convenient basis for the invariants of the group corresponding to $\mbox{\bf A}_{2,2}$
in the form
$$
I_1 = { \Delta x \over x}  ,
\qquad
I_2 = {  \Delta y    \over   \Delta   x  }   ,
\qquad
I_3 =  \dot{y}     .
$$
The general invariant DODE with delay can be written as
\begin{equation}       \label{DODEcase23}
\dot{y}   =    f  \left( {  \Delta y    \over     \Delta   x   }  \right) ,
\qquad
   x_-    =    x   g  \left( {  \Delta y    \over     \Delta   x   }  \right)   .
\end{equation}

\bigskip


$\mbox{\bf A}_{2,3}$:
The Abelian Lie algebra $2 {\sso {n}}_{1,1}$  with connected basis elements
\begin{equation}      \label{Dcase22}
X_1 = { \ddy}   ,
\qquad
X_2 = x  { \ddy} .
\end{equation}
A basis for the  invariants of the group corresponding to the Lie algebra $\mbox{\bf A}_{2,3}$ is
$$
I_1 = x ,
\qquad
I_2 = x_- ,
\qquad
I_3 =  \dot{y}   -   {  \Delta y    \over   \Delta  x  }
$$
and the invariant DODE with equation for delay can be presented as
\begin{equation}       \label{DODEcase22}
\dot{y}  =   {  \Delta y    \over   \Delta  x  }    +  f (x)   ,
\qquad
   x_-    =     g ( x )   .
\end{equation}
We shall assume that $ f(x) {\not\equiv} 0 $;
otherwise~(\ref{DODEcase22}) is a special case of~(\ref{DODEcase24}).

\bigskip


$\mbox{\bf A}_{2,4}$:
The Abelian Lie algebra $2 {\sso {n}}_{1,1}$ with non{connected} basis elements
\begin{equation}      \label{Dcase21}
X_1 = { \ddx}  ,
\qquad
X_2 = { \ddy}   .
\end{equation}
A convenient set of invariants is
$$
I_1 = \Delta  x  ,
\qquad
I_2 = \Delta  y ,
\qquad
I_3 = \dot{y}     .
$$
The most general invariant DODE and delay relation can be written as
\begin{equation}     \label{DODEcase21}
 \dot{y}  =  f  \left(  {  \Delta y    \over  \Delta x  }  \right)  ,
\qquad
 \Delta   x = g (  \Delta y  )     .
\end{equation}

Using the classification of first-order DODEs admitting  two-dimensional symmetry algebras,
we can formulate following general results.

\begin{theorem}  \label{lineartheorem}
Let the DODE~(\ref{DODE}) with delay relation~(\ref{delay})
 admit two linearly connected symmetries.
Then it can be transformed into
the homogeneous linear equation~(\ref{DODEcase24})
if the symmetries do not commute;
and to  the inhomogeneous linear equation~(\ref{DODEcase22})
if these symmetries commute.

In both cases the DODE is supplemented by the delay relation $ x_- = g(x) $,
which does not depend on the solutions of the considered DODE.
\end{theorem}

\noindent {\it Proof.} If there are two linearly connected
symmetries, they can be transformed
into the form~(\ref{Dcase24}) in non-Abelian case
and
into the form~(\ref{Dcase22}) in Abelian case.
The transformation or a special case of the transformation will transform
the corresponding DODE into equation~(\ref{DODEcase24}) or~(\ref{DODEcase22}), respectively.

The delay equation will be transformed into $ x_- = g(x) $ in both cases.
\hfill $\Box$

\medskip

\noindent {\bf Comment.}
{\it From now on we shall use the term linear DODS
precisely for the systems~(\ref{DODEcase24}) and~(\ref{DODEcase22}).
}

\medskip

Thus, DODEs which admit two linearly connected symmetries are linearizable.

\begin{corollary}
Any Lie algebra containing a two-dimensional sub{algebra}
realized by linearly connected vector fields will provide
invariant DODS that can be transformed into the form
(\ref{DODEcase24}) or~(\ref{DODEcase22})
(i.e. linear DODE with a solution independent delay relation).
The larger Lie algebra will at most put constraints
on the functions $f(x)$ and $g(x)$.
\end{corollary}

For $ n = \mbox{dim} \ L \geq 3 $ we will consider only nonlinear DODSs in this section.
Linear ones are considered in Section~\ref{linearsection}.

\subsection{Dimension 3. Nonlinear DODS.}

\label{Nonlinear_DODS}

From Table~1 we see that we have 15 classes of algebras to consider.
Among them
$\mbox{\bf A}_{3,1}$, $\mbox{\bf A}_{3,3} ^a $, $\mbox{\bf A}_{3,5}$,
$\mbox{\bf A}_{3,7} ^b $, $\mbox{\bf A}_{3,11}$, $\mbox{\bf A}_{3,13}$,
$\mbox{\bf A}_{3,14}$ and $\mbox{\bf A}_{3,15}$
contain the sub{algebras}
$\mbox{\bf A}_{2,1}$ or $\mbox{\bf A}_{2,3}$
and hence lead to linear DODSs.
Here we will treat the remaining ones.

\subsubsection{Specifications of   DODS~(\ref{DODEcase21}):}


$\mbox{\bf A}_{3,2}  ^a $:
The Lie algebra ${\sso {s}}_{3,1}$ can be realized as
\begin{equation}           \label{Dcase39}
X_1 =  { \ddx} ,
\qquad
X_2 =  { \ddy} ,
\qquad
X_3 =   x  { \ddx}  +  a  y  { \ddy}  ,
\qquad
 0 < |a| \leq 1 .
\end{equation}
We consider two sub{cases}:

\begin{itemize}

 \item

 $ a \neq 1 $


A basis for invariants is given by
$$
I_1 =    {  \Delta y  \over ( \Delta x )  ^a  }       ,
\qquad
I_2 =    \dot{y}   {    \Delta x    \over    \Delta y   }    .
$$
All invariant DODSs are given by
\begin{equation}      \label{DODEcase39}
 \dot{y}  =   C_1  { \Delta y  \over \Delta x  } ,
\qquad
  \Delta  x      =     ( \Delta y )  ^{ 1/a }     ,
\qquad
 C_1 \neq 0 .
\end{equation}

It is interesting to note that in~(\ref{DODEcase39})
the DODE is actually linear,
but the delay relation is solution dependent and nonlinear.

 \item

$ a = 1 $



A basis for the invariants is
$$
I_1 =   { \Delta y  \over \Delta x  }    ,
\qquad
I_2 =     \dot{y}   .
$$
In this case there is no invariant DODE~(\ref{DODE}) with invariant delay relation~(\ref{delay}).
The obtained invariant equations are
\begin{equation}           \label{attempt}
     \dot{y}  = f \left(    { \Delta y  \over \Delta x  }      \right) ,
\qquad
{ \Delta y  \over \Delta x }= C \neq 0  .
\end{equation}
This implies $ \dot{y}  = f  (    C  )  $, a trivial ODE, rather than a DODE.


\end{itemize}

\bigskip


$\mbox{\bf A}_{3,4}$:
The solvable Lie algebra ${\sso {s}}_{3,2}$ can be represented as
\begin{equation}          \label{Dcase38}
X_1 = { \ddx} ,
\qquad
X_2 =  { \ddy} ,
\qquad
X_3 =   x  { \ddx}  +   ( x  +  y )  { \ddy}
\end{equation}
with $X_1$ and $X_2$ linearly non{connected}.
A basis for the  invariants is
$$
I_1 =    { e^ { \Delta y  \over \Delta x  }     \over \Delta x }     ,
\qquad
I_2 =      \dot{y}   -  { \Delta y  \over \Delta x  }
$$
and the general invariant DODE with the invariant delay relation can be written as
\begin{equation}      \label{DODEcase38}
 \dot{y}  =   { \Delta y  \over \Delta x  }    +   C_1 ,
\qquad
 \Delta  x    =  C_2   e^ { \Delta y  \over \Delta x  }   ,
\qquad
C_2  \neq 0  .
\end{equation}
In~(\ref{DODEcase38}) the DODE is linear,
however the delay relation is nonlinear for $  \Delta x $
and depends on $ \Delta y $.

\bigskip


$\mbox{\bf A}_{3,6} ^b$:
The algebra ${\sso {s}}_{3,3}$ can be represented as
\begin{equation}           \label{Dcase310}
X_1 =  { \ddx} ,
\qquad
X_2 =  { \ddy} ,
\qquad
X_3 =   (b x + y )  { \ddx}  +  ( b  y - x )   { \ddy}  ,
\quad
b \geq 0 .
\end{equation}
Over $\mathbb{C}$  this algebra is equivalent to $\mbox{\bf A}_{3,2} ^a $.
The corresponding transformation of variables is
$$
t = { x - i y \over \sqrt{2} } ,
\qquad
 u  =  -  { i \over \sqrt{2}} ( x + i y )   .
$$
A basis for the invariants can be chosen in the form
$$
I_1 =    \Delta x     e^{ b   \arctan    {  \Delta y  \over  \Delta x   }   }
\sqrt{  1 + \left( {  \Delta y  \over  \Delta x   }  \right) ^2  }      ,
\qquad
I_2 =
 { \dot{y}    -  {\displaystyle  {  \Delta y  \over  \Delta x   }    }
\over
1 +    \dot{y}    {\displaystyle    {  \Delta y  \over  \Delta x   }  }    }  .
$$
The general form of the invariant DODE with the invariant delay relation are
\begin{equation}      \label{DODEcase310}
   \dot{y}
= {  {\displaystyle  {  \Delta y  \over  \Delta x   }  }    + C_1
 \over
1  -   C_1    {\displaystyle  {  \Delta y  \over  \Delta x   }  }  } ,
\qquad
 \Delta x     e^{ b   \arctan    {  \Delta y  \over  \Delta x   }   }
\sqrt{  1 + \left( {  \Delta y  \over  \Delta x   }  \right) ^2 }  = C_2     ,
\qquad
C_2  \neq 0  .
\end{equation}

\subsubsection{Specifications  of     DODS~(\ref{DODEcase23}):}


$\mbox{\bf A}_{3,8}$:
This is the first of four inequivalent realizations of $   {\sso {sl}}  (2, \mathbb{F})$.
We have
\begin{equation}         \label{Dcase311C}
X_1 = { \ddy} ,
\qquad
X_2 =  x { \ddx} +  y   { \ddy} ,
\qquad
X_3 =   2 x y  { \ddx}  +   y ^2  { \ddy}  .
\end{equation}
A set of invariants
$$
I_1 =  { ( \Delta y )^2   \over   x  x_- }        ,
\qquad
I_2 =     { 1 \over  \dot{y}  }   -   {  2  x  \over  \Delta y   }
$$
provides us with invariant DODE and  the invariant delay relation
\begin{equation}       \label{DODEcase311C}
  \dot{y}     =  {  \Delta y   \over 2 x + C_1  \Delta y   } ,
\qquad
    x x_-     =  C_2    (\Delta y)^2   ,
\qquad
C_2  \neq 0   .
\end{equation}

We mention that this particular realization of $   {\sso {sl}}  (2, \mathbb{F})$
is not maximal in $ \mbox{diff} (2, \mathbb{F})$.
Indeed, we have an embedding
$$
 {\sso {sl}}   (2, \mathbb{F})
\subset
 {\sso {gl}}  (2, \mathbb{F})
\subset
 {\sso {sl}}  (3, \mathbb{F})
\subset
\mbox{diff} (2, \mathbb{F})
$$
and the centralizer of $\mbox{\bf A}_{3,8}$ is $ Y = x  ( \partial / \partial x )   $.
Note also that the coefficients of $ \partial / \partial y $ in $X_i$ depend
only on the variable y.
The realization is hence imprimitive.

\begin{remark}   (Alternative version)
This Lie algebra is often (and equivalently) realized by the vector fields
\begin{equation}          \label{Dcase311A}
X_1 =   { \ddx} ,
\qquad
X_2 =  2 x  { \ddx}  +  y  { \ddy} ,
\qquad
X_3 =   x^2   { \ddx}  +   x y  { \ddy}  .
\end{equation}
There are two invariants
$$
I_1 =  {  \Delta x   \over   y y_- }        ,
\qquad
I_2 =  y \left(    \dot{y}    -   {  \Delta y  \over  \Delta x   }    \right)    ,
$$
which provide us with the invariant DODS
\begin{equation}     \label{DODEcase311A}
    \dot{y}    = {  \Delta y  \over  \Delta x   }       +   {  C_1   \over y  } ,
\qquad
 \Delta x     =  C_2     y y_-   ,
\qquad
C_2  \neq 0  .
\end{equation}
\end{remark}



\bigskip


$\mbox{\bf A}_{3,9}$:
A second realization of  $  {\sso {sl}}   (2, \mathbb{F})$ is
\begin{equation}          \label{Dcase312B}
X_1 = { \ddy}  ,
\qquad
X_2 =  x  { \ddx}  +  y  { \ddy}  ,
\qquad
X_3 =   2 x y  { \ddx}  +   ( y ^2  - x^2 )  { \ddy}   .
\end{equation}
The  invariants can be chosen in the form
$$
I_1 =  { ( \Delta y )^2  + ( x  - x_- )^2  \over   x x_- }        ,
\qquad
I_2 =
 {    2 \Delta y  x   \dot{y}    -       ( \Delta y )^2  -  x_- ^2  +  x ^2
\over
 2 \Delta y  x   +      (    ( \Delta y )^2  + x_- ^2  - x ^2    )      \dot{y}  }  .
$$
We write the invariant DODE with the delay relation as
\begin{equation}     \label{DODEcase312B}
      \dot{y}
= {       { ( \Delta y )^2  + x_- ^2  - x ^2 } + 2 C_1 x \Delta y
\over
2x \Delta y   - C_1  (  { ( \Delta y )^2  + x_- ^2  - x ^2 }  )  } ,
\qquad
{ ( \Delta y )^2  + ( x  - x_- )^2  }       =  C_2      x x_-     .
\end{equation}

The sub{algebras} $\mbox{\bf A}_{3,8}$ and $\mbox{\bf A}_{3,9}$ are not equivalent,
neither for $\mathbb{F} = \mathbb{C}$, nor for $\mathbb{F} = \mathbb{R}$.
To see this, it is sufficient to notice that there is no nonzero element of  $ \mbox{diff} (2, \mathbb{F})$
that commutes with all elements $X_i$  from~(\ref{Dcase312B}),
i.e., the centralizer of  $\mbox{\bf A}_{3,9}$    in   $ \mbox{diff} (2, \mathbb{F})$   is zero,
whereas that of   $\mbox{\bf A}_{3,8}$     is $ Y = x  (  \partial / \partial x )   $.
Over $ \mathbb{R}$ the algebra~(\ref{Dcase312B}) is primitive.
However, over $ \mathbb{C}$ we can put
$$
u = i x  ,
\qquad
v = y .
$$
The algebra~(\ref{Dcase312B})  in terms of the coordinates $u$ and $v$ is transformed into
the algebra $\mbox{\bf A}_{3,10}$,
which we turn to now.

\bigskip


$\mbox{\bf A}_{3,10}$:
A third realization of  $  {\sso {sl}}   (2, \mathbb{F})$ is
\begin{equation}         \label{Dcase313B}
X_1 = { \ddy} ,
\qquad
X_2 =  x  { \ddx}  +  y  { \ddy} ,
\qquad
X_3 =   2 x y  { \ddx}  +   ( y ^2  + x^2 )   { \ddy}   .
\end{equation}

Over $ \mathbb{R}$,
$\mbox{\bf A}_{3,10}$ is a new inequivalent imprimitive realization of   $  {\sso {sl}}   (2, \mathbb{F})$.
As previously mentioned $\mbox{\bf A}_{3,10}$
and  $\mbox{\bf A}_{3,9}$
are equivalent over $\mathbb{C}$.
The realization is imprimitive.

A complete set of invariants is
$$
I_1 =  { ( \Delta y )^2  - ( x  - x_- )^2  \over   x x_- }        ,
\qquad
I_2 =
 {   2 \Delta y x  \dot{y}    -      ( \Delta y  )^2  - x^2  +  x_- ^2
\over
 2 \Delta y x    -        (      ( \Delta y  )^2  + x^2 - x_- ^2    )     \dot{y}    } .
$$
The invariant DODE with the delay relation an be written  as
\begin{equation}      \label{DODEcase313B}
   \dot{y}
= {  {  ( \Delta y  )^2  + x^2 - x_- ^2  }      + 2 C_1 x  \Delta y
\over
 2 x  \Delta y  + C_1 ( {  ( \Delta y  )^2  + x^2 - x_- ^2  }   ) } ,
\qquad
{ ( \Delta y )^2  - ( x  - x_- )^2   }       =  C_2     x x_-      .
\end{equation}

We mention that $\mbox{\bf A}_{3,10}$ is not maximal in $ \mbox{diff} (2, \mathbb{F})$.
For $\mathbb{F} = \mathbb{R}$ and $\mathbb{F} = \mathbb{C}$,
we have
$$
\mbox{\bf A}_{3,10}
\subset
 {\sso {o}}   (2, 2)
\subset
\mbox{diff} (2, \mathbb{R})  ,
$$
or
$$
\mbox{\bf A}_{3,10}
\subset
 {\sso {o}}   (4, \mathbb{C} )
\subset
\mbox{diff} (2, \mathbb{C})  ,
$$
respectively.
The algebras $  {\sso {o}}   (2, 2) $
and $  {\sso {o}}   (4, \mathbb{C} ) $ are both realized by the vector fields
$$
\left[
{ \ddx} ,
\quad
x { \ddx} ,
\quad
x^2 { \ddx} ,
\quad
{ \ddy}  ,
\quad
y  { \ddy} ,
\quad
y^2  { \ddy}
\right] .
$$

\begin{remark}   (Alternative version) For this Lie algebra
we can use another realization
\begin{equation}        \label{Dcase313C}
X_1 = { \ddx} +  { \ddy} ,
\qquad
X_2 = x  { \ddx}  +  y { \ddy} ,
\qquad
X_3 = x^2  { \ddx} +  y^2  { \ddy}  .
\end{equation}
We obtain  invariants
$$
I_1 = { ( y - y _- ) ( x - x_- )  \over   ( y - x ) ( y _- - x _- ) } ,
\qquad
I_2 = \left( {  x - x_-   \over   y - x_- }  \right)  ^2      \dot{y}
$$
and the invariant DODS
\begin{equation}     \label{DODEcase313C}
\dot{y}  =  C_1  \left(  { y - x_- \over  x - x_- } \right) ^2 ,
\qquad
{ ( y - y _- ) ( x - x_- )  \over   ( y - x ) ( y _- - x _- ) } = C_2   ,
\qquad
C_1 C_2  \neq 0  .
\end{equation}
\end{remark}

\subsubsection{Other DODS (not a specification a DODS given by a two-dimensional Lie algebra) }


$\mbox{\bf A}_{3,12}$:
There exists just one (up to equivalence) realization of $ {\sso {o}} (3)$ as a sub{algebra} of
$\mbox{diff} (2, \mathbb{F}) $.
We choose it in the form
\begin{multline}          \label{Dcase314}
X_1 = ( 1 + x^2  )  { \ddx} + xy  { \ddy} ,
\qquad
X_2 = xy  { \ddx} + ( 1 + y^2  )  { \ddy} ,
\qquad
X_3 = y  { \ddx}  -  x  { \ddy}  .
\end{multline}
The invariants can be chosen to be
$$
I_1 = {
 ( x - x _- )^2
\left( 1  + \left( { \displaystyle { y - y _- \over x - x_- }  } \right)^2
+ \left( y - x { \displaystyle {  y - y _- \over  x - x_- } } \right)^2   \right)
\over
(  1 + x^2 +    y ^2  )
(  1 + x _- ^2 +    y _- ^2  )
}  ,
$$
$$
I_2 = {
 ( x - x _- )
\left( \dot{y} -  { \displaystyle { y - y _- \over x - x_- } } \right)
\over
\sqrt{  1 +  \dot{y} ^2 +  ( y - x  \dot{y} ) ^2  }
\sqrt{  1 + x _- ^2 +    y _- ^2  }
}  .
$$
The general form of the invariant DODE with delay relation can be written as
\begin{equation*}
 {
 ( x - x _- )
\left( \dot{y} -  { \displaystyle { y - y _- \over x - x_- } } \right)
\over
\sqrt{  1 +  \dot{y} ^2 +  ( y - x  \dot{y} ) ^2  }
\sqrt{  1 + x _- ^2 +    y _- ^2  }
}
= C_1 ,
\end{equation*}
\begin{equation}    \label{DODEcase314}
{
 ( x - x _- )^2
\left( 1  + \left( { \displaystyle { y - y _- \over x - x_- }  } \right)^2
+ \left( y - x { \displaystyle {  y - y _- \over  x - x_- } } \right)^2   \right)
\over
(  1 + x^2 +    y ^2  )
(  1 + x _- ^2 +    y _- ^2  )
}
= C_2  ,
\qquad
C_2  \neq  0   .
\end{equation}

Over $\mathbb{C} $ the algebra $\mbox{\bf A}_{3,12}$
is equivalent to the  $  {\sso {sl}}  ( 2, \mathbb{R} ) $  algebra  $\mbox{\bf A}_{3,10}$.
The transformation of variables that takes one algebra into the other one is quite complicated and we
shall not reproduce it here.
The algebra $\mbox{\bf A}_{3,12}$  does not have a nontrivial centralizer in $\mbox{diff} (2, \mathbb{F}) $.

\bigskip

\noindent { \bf We conclude this point with the following observations: }

\begin{itemize}



\item

For $\mbox{\bf A}_{3,2} ^a $ with $ a \neq 1 $  and $\mbox{\bf A}_{3,4}$
we get linear DODEs. However, the delays are solution dependent.

\item

For  $\mbox{\bf A}_{3,2} ^a $  with $ a = 1 $
there is no invariant DODS.

\item

For   $\mbox{\bf A}_{3,6}^b$,   $\mbox{\bf A}_{3,8}$,  $\mbox{\bf A}_{3,9}$,
$\mbox{\bf A}_{3,10}$  and   $\mbox{\bf A}_{3,12}$
we generally get nonlinear DODEs with nonlinear delay relations.

\item

Contrary to the case of $ \mbox{dim} \ L < 3 $,
no  arbitrary functions appear in the invariant DODS,
only arbitrary constants.

\end{itemize}

\subsection{Dimensions $n \geq 4$}

In this subsection we shall show that Lie algebras of dimension
$ n \geq 4 $ can only lead to linear DODEs with $y$-independent delay relations.
We again start from Table~1. All{together} 13 isomorphism classes of
real four-dimensional indecomposable Lie algebras exist~\cite{SWbook}
as well as 9 classes of decomposable ones.
Some of them cannot be realized by vector fields,
others are realized one or two times.
Finally, Table~1 contains a list of 22 realizations.

Among them  $\mbox{\bf A}_{4,1}$, ..., $\mbox{\bf A}_{4,7}$,
$\mbox{\bf A}_{4,12}$, $\mbox{\bf A}_{4,14}$, ..., $\mbox{\bf A}_{4,18}$
and   $\mbox{\bf A}_{4,21}$ contain 3-dimensional linearly connected sub{algebras};
$\mbox{\bf A}_{4,8}$, ..., $\mbox{\bf A}_{4,11}$,   $\mbox{\bf A}_{4,19}$ and  $\mbox{\bf A}_{4,20}$
contain 2-dimensional linearly connected sub{algebras}.
In the Abelian algebra   $\mbox{\bf A}_{4,22}$  all 4 vector fields are linearly connected.
All these algebras (all four-dimensional algebras except    $\mbox{\bf A}_{4,13}$)
will hence lead to invariant linear DODEs with solution independent
delay relations.



For $ n \geq 5 $ only one algebra in the list of Ref.~\cite{Gonzalez1992}
does not contain a two-dimensional linearly connected sub{algebra}, namely, the
6-dimensional algebra $ {\sso {so}} (3,1)$:
$$
X_1 =   { \ddx}  ,
\qquad
X_2 =   { \ddy}  ,
\qquad
X_3 =  x { \ddx}  + y { \ddy}  ,
\qquad
X_4 =  y { \ddx}  - x { \ddy}  ,
$$
$$
X_5 =  (x^2 - y^2)  { \ddx}  + 2 x y { \ddy}  ,
\qquad
X_6 =  2 x y { \ddx}  + ( y^2 - x^2 ) { \ddy}  .
$$
From Table~1 we see that
$ {\sso {so}} (3,1) $ contains $ \mbox{\bf A}_{4,13}  $ as a sub{algebra}.

It remains to show that no DODS invariant under $ \mbox{\bf A}_{4,13}  $
exists (this also rules out $ {\sso {so}} (3,1) $):
For $ \mbox{\bf A}_{4,13}  $ there is only one invariant
$$
I_1 =   {  \displaystyle  \dot{y}  -  { y - y _- \over x - x _- }
\over
\displaystyle   1 +  \dot{y}   { y - y _- \over x - x _- }  }  .
$$
It can be used to construct an invariant  DODE as $I_1 = C_1$,
but it is not possible to supplement
this DODE with an invariant delay relation.
We arrive at the following Theorem.

\begin{theorem}
The only DODSs of the form~(\ref{DODE}),(\ref{delay})
that have symmetry algebras of dimension
$n \geq 4$ are equivalent to linear DODSs with solution independent delay equations.
\end{theorem}

Linear DODSs will be treated separately in Section~\ref{linearsection}.

\section{Linear DODSs}

\label{linearsection}

In the previous section we observed that two-dimensional Lie algebras of linearly connected
vector fields provide us with linear DODEs  and solution independent delay relations.
Two DODSs of this type were obtained:~(\ref{DODEcase24})    and~(\ref{DODEcase22}).
We remark that {\it method of steps} allows us to solve these systems using successive quadratures.

Let us show that they actually admit infinite-dimensional symmetry algebras
due to the linear superposition principle.
It is convenient to consider the most  general linear DODS
in the considered class, namely
\begin{equation}    \label{linearDODE}
 \dot{y} = \alpha  ( x  )  y +   \beta  ( x  )  y_-  +  \gamma  ( x  )  ,
\qquad
x_-   = g  ( x )   ,
\end{equation}
where $\alpha (x)$, $ \beta (x) $,  $ \gamma (x) $ and $ g(x) $
are arbitrary real functions,
smooth in some interval $ x \in I $, satisfying
\begin{equation}
\beta (x)  {\not\equiv} 0 ,
\qquad
g (x)  <  x,
\qquad
g (x)  {\not\equiv}  \mbox{const}  .
\end{equation}
We will also need its homogeneous  counterpart
\begin{equation}   \label{homogeneous}
\dot{y}  =  \alpha  ( x  )   y   +   \beta   ( x  )   y _-  ,
\qquad
x_-   = g  ( x )    .
\end{equation}
Note that equation~(\ref{linearDODE}) includes~(\ref{DODEcase24})    and~(\ref{DODEcase22})
as particular cases.

\begin{proposition}    \label{homochange}
The change of variables
\begin{equation}
\bar{x} = x,
\qquad
\bar{y} = y - \sigma (x),
\end{equation}
where  $ \sigma (x)$ is an arbitrary solution of the inhomogeneous DODS~(\ref{linearDODE}),
transforms the inhomogeneous DODS~(\ref{linearDODE})
into its  homogeneous  counterpart~(\ref{homogeneous}).
\end{proposition}

\begin{theorem}    \label{th41a}
Consider the linear DODS~(\ref{linearDODE}).
For all functions $ \alpha (x) $, $ \beta (x) $,  $ \gamma (x) $ and $ g (x) $
the DODS admits an infinite-dimensional symmetry algebra
represented by the vector fields
\begin{equation}    \label{always}
X (\rho)  =  \rho  (x)  { \ddy} ,
\qquad
Y (\sigma)  =  ( y  -    \sigma  (x) )  { \ddy}  ,
\end{equation}
where $ \rho (x) $  is the general solution of the homogeneous DODS~(\ref{homogeneous})
and  $ \sigma  (x) $  is any one particular solution of the DODS~(\ref{linearDODE}).
\end{theorem}

\noindent {\it Proof.}
Application of the symmetry operator
\begin{equation*}
X = \xi (x,y) { \ddx} +  \eta (x,y) { \ddy}
\end{equation*}
to the delay equation in~(\ref{linearDODE}) gives
\begin{equation}
 \xi (x_- ,y_- ) = \dot{g} (x)  \xi (x,y)  .
\end{equation}
Since $x$, $y$ and $y_-$ can be considered as independent
while $x_-$ and $\dot{y}$ are related to them via Eqs.~(\ref{linearDODE})
we get
\begin{equation}   \label{conncetion1}
\xi = \xi(x), \qquad  \xi ( g ( x )  ) = \dot{g} (x)  \xi (x) .
\end{equation}
Now we consider the application of the symmetry to the DODE in~(\ref{linearDODE}) on the solutions of this equation:
\begin{multline}   \label{deter2}
\eta  _x  (x,y) + ( \eta _y (x,y) - \dot{\xi} (x) )    ( \alpha  ( x )  y +   \beta  ( x )  y_-  +  \gamma  ( x )  )  \\
=   \xi(x) (    \dot{\alpha}  ( x )  y +    \dot{\beta}  ( x )  y_-  +   \dot{\gamma}  ( x )  )
+  \alpha  ( x )   \eta  (x,y)  +   \beta  ( x )  \eta  ( x_- , y_-)    .
\end{multline}
Differentiating with respect to $y_-$ twice, we obtain $ \eta _{ y_- y_- } (x_- , y_-)  = 0 $,
i.e.
\begin{equation}
\eta (x,y) = A (x) y + B (x)  .
\end{equation}
Substituting this result into equation~(\ref{deter2}),
we get the system
\begin{equation}   \label{split1}
 \dot{A} (x)
=   \alpha (x)   \dot{\xi} (x)    +    \dot{\alpha} (x)   \xi(x)    ,
\end{equation}
\begin{equation}    \label{split2}
\beta (x)  \dot{\xi} (x)    +   \dot{\beta} (x)   \xi(x)
=   \beta (x) (  A (x)   -  A(x _-)     )   ,
\end{equation}
\begin{equation}    \label{split3}
 \dot{B} (x)  =   \alpha (x) B(x) + \beta (x) B(x_-)
+  \gamma  (x)   \dot{\xi} (x)      +  \dot{\gamma} (x)   \xi(x)  -  \gamma  (x)  A (x)   .
\end{equation}

Equation~(\ref{split1}) can be integrated as
\begin{equation}     \label{solutionA}
A (x) =     \xi (x)  \alpha (x)  + A_0  ,
\qquad
A_0 = \mbox{const} .
\end{equation}
Substitution of this result into the equation~(\ref{split2})  gives
\begin{equation}   \label{condsss}
\dot{\xi}(x) =
\left[
 \alpha (x) -  \dot{g} (x) \alpha (g(x)) - { \dot{\beta} (x) \over \beta (x) }
\right]
\xi(x)  .
\end{equation}
For general $\alpha (x) $, $\beta (x) $ and $g(x)$
equations~(\ref{conncetion1}) and~(\ref{condsss})
have only one solution, namely  $ \xi (x) \equiv  0 $.
Thus we obtain symmetries with coefficients
$$
\xi (x,y)   \equiv    0 ,
\qquad
\eta (x,y) =  A_0 y  + B(x)     ,
\qquad
A_0 = \mbox{const} ,
$$
where $ B(x) $ solves
\begin{equation*}
 \dot{B} (x)     =   \alpha (x) B(x) + \beta (x) B(x_-) - A_0 \gamma (x)   ,
\qquad
x_- = g (x) .
\end{equation*}
Finally, we can rewrite the admitted symmetries
as  given in~(\ref{always}).
\hfill $\Box$

\medskip

\begin{remark}
For the homogeneous linear DODS~(\ref{homogeneous})
the theorem gets simplified since we can use $ \sigma  (x)  = 0 $.
We get an infinite-dimensional symmetry algebra
represented by the vector fields
\begin{equation}    \label{homoalways}
 X  (\rho)  =  \rho  (x)  { \ddy} ,
\qquad
 Y   =   y    { \ddy}  ,
\end{equation}
where  $ \rho (x) $  is the general solution of the homogeneous DODS~(\ref{homogeneous}).
\end{remark}


Let us continue to investigate
the determining equations~(\ref{conncetion1}),(\ref{split1}),(\ref{split2}),(\ref{split3})
and find when
the overdetermined discrete-differential system~(\ref{conncetion1}) and~(\ref{condsss})
can have nontrivial solutions $\xi (x) {\not \equiv 0 } $.
For convenience we present the ODE~(\ref{condsss}) as
\begin{equation}     \label{K_modified}
\dot{\xi}(x) = K(x) \xi(x)  ,
\qquad
K(x) =
 \alpha (x) -  \dot{g} (x) \alpha (g(x)) - { \dot{\beta} (x) \over \beta (x) } .
\end{equation}

Differentiating~(\ref{conncetion1})  with respect to $x$,
we get
$$
 \dot{\xi} ( g ( x )  ) \dot{g} (x)  = \ddot{g} (x)  \xi (x)  + \dot{g} (x)  \dot{\xi} (x) .
$$
Using~(\ref{conncetion1}) and~(\ref{K_modified}),
we obtain
$$
K(g(x)) ( \dot{g} (x) )^2  \xi (x)  = \ddot{g} (x)  \xi (x)  + \dot{g} (x)  K(x)  \xi  (x)   .
$$
From this  equation it follows that
ether  compatibility condition~(\ref{compatibil}) given below is satisfied
or $ \xi (x)  \equiv 0$.
The compatibility condition is required for existence of nontrivial  solution $ \xi (x)  {\not  \equiv}  0  $
which solve  the overdetermined system~(\ref{conncetion1}) and~(\ref{condsss}).
Then $ A(x) $ and $ B(x) $ can be found
from~(\ref{solutionA}) and~(\ref{split3}), respectively.

Finally, the admitted symmetry algebra can be gives as symmetries
(\ref{always}) supplemented by one additional symmetry.
We present this result as the following theorem.



\begin{theorem}    \label{th41b}
Consider the linear DODS~(\ref{linearDODE}).
For specific choices of the arbitrary functions $ \alpha (x) $, $ \beta (x) $ and $ g  (x) $,
namely for functions satisfying the  {compatibility condition}
\begin{equation}   \label{compatibil}
K(g(x)) ( \dot{g} (x) ) ^ 2  =  \ddot{g} (x)  + K(x)  \dot{g} (x) ,
\end{equation}
where
\begin{equation}    \label{Kdefinition}
K(x) = \alpha (x)   -   \dot{g} (x)  \alpha ( g (x) )  - { \dot{\beta} (x)  \over \beta (x) } ,
\end{equation}
the symmetry  algebra is larger. It contains one additional basis element of the form
\begin{equation}   \label{additional}
Z = \xi (x) { \ddx} +  ( A (x) y + B (x) )  { \ddy} ,
\qquad
\xi (x) {\not \equiv } 0,
\end{equation}
where
\begin{equation}     \label{xisolution}
\xi ( x ) = e^{ \int K(x) dx }  ,
\qquad
A(x) = \xi ( x ) \alpha(x)
\end{equation}
and the function $  B (x) $ is a particular solution of the DODS
\begin{equation*}    
\dot{B} (x)  =   \alpha (x) B(x) + \beta (x) B(x_-)
 +  \gamma  (x)   \dot{\xi} (x)      +  \dot{\gamma} (x)   \xi(x)  -  \gamma  (x)  A (x)  ,
\qquad
x_- = g (x)  .
 \end{equation*}
\end{theorem}

\begin{remark}
Though  $ \xi ( x )  $ is given as a solution of the ODE~(\ref{condsss})
it must also satisfy the equation~(\ref{conncetion1})
provided that the compatibility condition~(\ref{compatibil})  is satisfied.
\end{remark}

\begin{remark}
For the homogeneous linear DODS~(\ref{homogeneous})
the theorem gets simplified and we can present the admitted symmetries  as
\begin{equation}    \label{addhomoalways}
X  (\rho)  =  \rho  (x)  { \ddy} ,
\qquad
Y    =   y    { \ddy}  ,
\qquad
Z = \xi (x) { \ddx} +   \xi (x)  \alpha (x)  y    { \ddy}  ,
\qquad
\xi (x) {\not \equiv } 0,
\end{equation}
where  $ \rho (x) $  is the general solution of the homogeneous DODS~(\ref{homogeneous})
and   $ \xi (x) $ is a nontrivial solution of the overdetermined system~(\ref{conncetion1}) and~(\ref{condsss}),
which is given in~(\ref{xisolution}).
\end{remark}

Using  the results of the previous  theorems,
we can present all symmetries of the homogeneous DODS~(\ref{DODEcase24})
as
\begin{equation}
 X (\rho)  =  \rho  (x)  { \ddy} ,
\qquad
Y   =  y    { \ddy}  ,
\end{equation}
where $ \rho  (x) $ is an arbitrary solution of the DODS~(\ref{DODEcase24}).
The symmetries of the inhomogeneous DODS~(\ref{DODEcase22})
can be presented as
\begin{equation}
 X (\rho)  =  \rho  (x)  { \ddy} ,
\qquad
Y  (\sigma)  =  ( y  -    \sigma  (x) )  { \ddy}  ,
\end{equation}
where  $ \sigma  (x) $  is a particular solution of the original inhomogeneous DODS~(\ref{DODEcase22})
and $ \rho (x) $  is the general solution of its homogeneous counterpart.

We can use the results established above to provide simplifications
of the linear DODS.

\begin{theorem}    \label{th42new}
If the linear DODS~(\ref{linearDODE}) has functions
$\alpha (x)$, $ \beta (x) $ and $ g(x) $
satisfying the compatibility condition~(\ref{compatibil}),
the DODS can be transformed into the representative form
\begin{equation}      \label{canonical1new}
\dot{y} = y_-   + h  (x)  ,
\qquad
x_- = x - C,
\qquad
C  >  0 ,
\end{equation}
which admits symmetries
\begin{equation}
 X  ( \rho )  =  \rho  (x)  { \ddy} ,
\qquad
 Y (\sigma )    =   ( y   - \sigma (x) )  { \ddy}  ,
\qquad
 Z   = { \ddx} +  B(x)  { \ddy}    ,
\end{equation}
where  $ \sigma  (x) $  is any one particular solution of the inhomogeneous  DODS~(\ref{canonical1new}),
$ \rho   (x) $  is  the general solution of the corresponding homogeneous DODS
and $ B(x) $ is a particular solution of the DODS
\begin{equation*}
\dot{B} (x)  = B ( x_- )   + \dot{h} (x)  ,
\qquad
x_- = x - C,
\qquad
C  >  0 .
\end{equation*}
\end{theorem}

\noindent {\it Proof.}
Let us consider a linear DODS~(\ref{linearDODE})
which admits a symmetry of the form~(\ref{additional}) with
$ \xi (x) {\not \equiv} 0 $.

Variable change
\begin{equation}
\bar{x} = \int   { 1 \over \xi (x) } dx ,
\qquad
\bar{y} = e^{ - \int \alpha (x) dx }   y
\end{equation}
brings the DODS~(\ref{linearDODE})
into the form
\begin{equation}
 \dot{y} = C_1   y_-  +   \tilde{\gamma} (x)    ,
\qquad
x_-   =  x  - C_2  ,
\qquad
C_1  \neq 0   ,
\qquad
C_2 > 0 .
\end{equation}
Finally,  we can rescale $x$  to obtain the system~(\ref{canonical1new}).
\hfill $\Box$
\medskip

Further simplification is possible if we know one particular solution
of the DODS and can bring the DODE into the homogeneous form.

\begin{corollary}    \label{th42}
If the linear homogeneous  DODS~(\ref{homogeneous}) has functions
$\alpha (x)$, $ \beta (x) $ and $ g(x) $
satisfying the compatibility condition~(\ref{compatibil}),
the DODS can be transformed into the representative form
\begin{equation}      \label{canonical1}
\dot{y} = y_- ,
\qquad
x_- = x - C,
\qquad
C  >  0 ,
\end{equation}
which admits symmetries
\begin{equation}
 X  ( \rho )  =  \rho  (x)  { \ddy} ,
\qquad
 Y   =   y    { \ddy}  ,
\qquad
 Z   = { \ddx}   ,
\end{equation}
where  $ \rho (x) $  is the general solution of the DODS~(\ref{canonical1}).
\end{corollary}




Let us note that  we could  choose the representative form
\begin{equation}    \label{simplehomolinearDODE_2}
 \dot{y} = C_1 y + C_2   y_-   ,
\qquad
x_-   =  x  - C  ,
\qquad
C_2  \neq 0   ,
\qquad
C > 0 .
\end{equation}
instead of the form~(\ref{canonical1}).

\begin{theorem}    \label{th43new}
If the linear DODS~(\ref{linearDODE}) has functions
$\alpha (x)$, $ \beta (x) $ and $ g(x) $
which do not satisfy the compatibility condition~(\ref{compatibil}),
the DODS can be transformed into the representative form
\begin{equation}      \label{canonical2new}
\dot{y} = f(x) y_- + h (x) ,
\qquad
x_- = x - C,
\qquad
\dot{f} (x) {\not \equiv} 0 ,
\qquad
C >  0
\end{equation}
or into the form
\begin{equation}      \label{canonical3new}
\dot{y} = y_-   + h (x) ,
\qquad
x_- = g (x) ,
\qquad
\ddot{g} (x) {\not \equiv} 0 ,
\qquad
g (x) < x  .
\end{equation}
\end{theorem}

\noindent {\it Proof.}
Let us consider a linear DODS~(\ref{linearDODE})
which does not possess an additional  symmetry of the form~(\ref{additional}).

Then we can employ the variable change
\begin{equation}
\bar{x} = x  ,
\qquad
\bar{y} = e^{ - \int \alpha (x) dx }   y
\end{equation}
to bring the DODS~(\ref{linearDODE})
into the form
\begin{equation}    \label{simplehomolinearDODE2new}
\dot{y} = \tilde{\beta} (x)   y_-  + \tilde{\gamma} (x)   ,
\qquad
x_-   =  g(x)   .
\end{equation}

Now there are two possibilities.
We can straighten the delay relation  and
transform the DODS~(\ref{simplehomolinearDODE2new}) into the form
\begin{equation}    \label{simplehomolinearDODE2anew}
 \dot{y} = f (x)   y_-   + h (x) ,
\qquad
x_-   =  x - C   ,
\qquad
C >  0  .
\end{equation}
Since this DODS cannot admit an additional symmetry of the form~(\ref{additional})
with $\xi (x) {\not \equiv} 0$
the compatibility condition~(\ref{compatibil}) is  not satisfied.
Thus, we get $ \dot{f} (x) {\not \equiv} 0 $.

Alternatively,  we can use  the variable change
\begin{equation}
\bar{x} = \int \tilde{\beta} (x)  dx  ,
\qquad
\bar{y} =    y
\end{equation}
to bring the  DODS~(\ref{simplehomolinearDODE2new})
into the form
\begin{equation}    \label{simplehomolinearDODE3new}
 \dot{y} =   y_-   + h (x) ,
\qquad
x_-   =  g(x)   .
\end{equation}
Since there can be no additional symmetry of the form~(\ref{additional})
with $\xi (x) {\not \equiv} 0$
the compatibility condition~(\ref{compatibil}) must not hold.
Thus, we get the restriction $ \ddot{g} (x) {\not \equiv} 0 $.

\hfill $\Box$
\medskip

\begin{corollary}    \label{th43}
If the linear homogeneous DODS~(\ref{homogeneous}) has functions
$\alpha (x)$, $ \beta (x) $ and $ g(x) $
which do not satisfy the compatibility condition~(\ref{compatibil}),
the DODS can be transformed into the representative form
\begin{equation}      \label{canonical2}
\dot{y} = f(x) y_- ,
\qquad
x_- = x - C,
\qquad
\dot{f} (x) {\not \equiv} 0 ,
\qquad
C >  0
\end{equation}
or into the form
\begin{equation}      \label{canonical3}
\dot{y} = y_- ,
\qquad
x_- = g (x) ,
\qquad
\ddot{g} (x) {\not \equiv} 0 ,
\qquad
g (x) < x  .
\end{equation}
\end{corollary}

\section{Exact solutions of the DODSs.}

\label{invariantsolutions1}

\subsection{Strategy for obtaining particular exact solutions}

\subsubsection{Find the symmetry algebra $L$}

\label{particularsolutions1}

We are given a general DODS of the form~(\ref{DODE}),(\ref{delay})
where $f$ and $g$ are given functions of the indicated arguments.
We apply the standard algorithm described in Section~\ref{generaltheory}
to find the symmetry algebra of the system.
Thus, we put
\begin{equation}   \label{E12}
E_1 = \dot{y} - f  (x, y, y_-) ,
\qquad
 E_2 = x_-  - g  (x, y, y_-)
 \end{equation}
and  $ X _{\alpha} $ as in~(\ref{operator1}),
$  \mbox{\bf pr} X  _{\alpha} $ as in~(\ref{prolongation})
and request
\begin{equation}       \label{determining}
\left.   \mbox{\bf pr} X  _{\alpha}  E_1   \right| _{ E_1 =0 , \   E_2 =0 } =  0  ,
\qquad
\left.   \mbox{\bf pr} X  _{\alpha}  E_2   \right| _{ E_1 =0 , \   E_2 =0 } =  0  .
 \end{equation}
Eqs.~(\ref{determining}) provide us with the determining equations for the coefficients
$\xi  _{\alpha} $  and  $\eta   _{\alpha} $ in~(\ref{operator1}).
From the analysis in Sections~\ref{Classification} and~\ref{linearsection}
it follows that the possible dimensions of the symmetry algebra $L$ are
$ n = 0, \ 1, \ 2, \ 3 $ and $ n = \infty $.

\subsubsection{Finite-dimensional symmetry algebras}

\label{particularsolutions3}

It follows from Sections~\ref{Classification} and~\ref{linearsection}
that the symmetry algebra
is finite-dimensional only if the DODS is genuinely nonlinear,
i.e. cannot be linearized by a point transformation.
In this case we have    $ 0  \leq  n  \leq 3  $.
The following can be done for genuinely nonlinear DODSs:

\begin{itemize}

\item

$ n = 0 $

In this case group theory is of no use.

\item

$ n = 1 $

We have only one vector fields $X$  and transform it into  $X = \partial _y $.
This simplifies DODS to the form~(\ref{DODEcase11})
involving two functions of two variables only.
This is a great simplification but does not provide any exact solutions.

\item

$ n = 2 $

Since we are now only  interested in nonlinear DODSs
the vector fields $X_1$  and $X_2$ must be  linearly {non}connected.
We obtain algebras   $\mbox{\bf A}_{2,2} $ and $\mbox{\bf A}_{2,4} $
with DODSs~(\ref{DODEcase23}) and~(\ref{DODEcase21}), respectively.

\item

$ n = 3 $

We obtain all the algebras and invariant DODSs of Section~\ref{Nonlinear_DODS}.

\end{itemize}

\subsection{Invariant solutions}

\subsubsection{General method}

If the DODS~(\ref{DODE}),(\ref{delay})
has a symmetry algebra $L$ of dimension $ n = \mbox{dim} \   L  \geq 1$,
it can, at least in principle, be used to construct explicit  analytical solutions.
These particular solutions will satisfy very specific initial conditions.
Essentially the method consists of construction solutions that are invariant
under a subgroup of the symmetry  group of the DODS.

In the case of the DODS~(\ref{DODE}),(\ref{delay})
it is sufficient to consider one-dimensional {sub}algebras.
They will all have the form
\begin{equation}
X = \sum  _{ \alpha = 1 } ^n   c _{ \alpha}   X _{ \alpha} ,
 \qquad
  c _{ \alpha}  \in \mathbb{R}  ,
\end{equation}
where   $c _{ \alpha} $ are constants and $  X _{ \alpha}  $ are of the form~(\ref{operator1})
and are elements of the symmetry algebra $L$.

\bigskip

\noindent   The method consists of several steps.

\begin{enumerate}

\item

Construct a representative list of one-dimensional {sub}algebras of   $L_i $
of the symmetry algebra $L$ of the DODS.
The  {sub}algebras are classified under the group of inner automorphisms $ G  = \mbox{exp} L$.

\item

For each {sub}algebra in the list calculate the invariants of the subgroups    $ G_i   = \mbox{exp} L_i $
in the four-dimensional space with local coordinates $ \{  x, y, x_- , y_- \} $.
There will be three functionally independent invariants.
For the method to be applicable two of the invariants must depend on two variables only,
namely  $(x,y) $ and   $(x_- ,y_- ) $ respectively. We set them equal to constants as follows:
\begin{equation}   \label{invariantsform}
J_1 (x,y) = A,
\qquad
J_2  ( x, y, x_- , y_- )  = B .
\end{equation}
They  must satisfy the Jacobian condition
\begin{equation}
\mbox{det}   \left(  { \partial ( J_1 , J_2 )  \over  \partial ( y  , x _- ) }  \right)     \neq   0
\end{equation}
 (we have  $ J_3  (  x_- , y_- )  =   J_1  ( x_- , y_- )  $,
i.e.,    $ J_3  (  x_- , y_- )  $ is obtained by shifting  $ J_1  (  x, y)  $ to $ (  x_- , y_- ) $).

All elements of the Lie algebra have the form~(\ref{operator1}).
A {\bf necessary condition} for invariants of the form~(\ref{invariantsform}) to exist
is that at least one of the vector fields in $L_i$  satisfies
\begin{equation}   \label{condition_for_solution}
\xi (x,y) {\not \equiv} 0 .
\end{equation}

\item

Solve equations~(\ref{invariantsform})  to obtain the reduction formulas
\begin{equation}  \label{reduction_formulas}
y = h (x, A) ,
\qquad
x_-  = k ( x, A, B)
\end{equation}
(we also have $y_-  = h (x_- , A)$).

\item

Substitute the reduction formulas~(\ref{reduction_formulas})
into the DODS~(\ref{DODE}),(\ref{delay})    and request that the equations should be satisfied  identically.
This will provide relations which define the constants
and therefore determine the functions $h$  and $k$.
It may also impose constraints on the functions $ f ( x, y, y_-)$  and $ g ( x, y, y_-)$
in~(\ref{DODE}) and~(\ref{delay}).
Once the relations are satisfied the invariant solution is given by~(\ref{reduction_formulas}).

\item

Apply the entire group $G_i = \mbox{exp} L_i $ to the obtained invariant solutions.
This can provide a more general solution depending on up to
$ ( \mbox{dim} \ L - 1 )$  parameters
(corresponding to the factor algebra $ L / L_i$).
Check whether the DODS imposes further constraints involving the new-parameters.

\end{enumerate}

Following Olver's~\cite{Olver1986} definition for ODEs and PDEs,
we  shall call solutions obtained in this manner
{\it group invariant solutions } (for DODSs).
These solutions are solutions of DODSs supplemented by appropriate initial conditions.

\subsubsection{Solutions of nonlinear DODSs}

In  this subsection we will apply the algorithm described above to all nonlinear systems 
obtained in Section~\ref{Classification}.
We only consider   symmetry algebras containing at least one vector field satisfying
$ \xi (x,y) {\not \equiv} 0 $.
This rules out the algebra   $ \mbox{\bf A}_{1,1} $.

To obtain representative list of all {sub}algebras   of the symmetry algebras
we can use the tables given in~\cite{PateraWinternitz}
(after appropriately identifying basis elements).
Alternatively one can calculate them in each case using the method presented in the same article.

As shown in Section~\ref{Classification},
for nonlinear DODSs the dimension satisfies $ \mbox{dim} \  L \leq 3$.
We will only present {sub}algebras that lead to invariant solutions.

\subsection{DODSs with $ \mbox{dim} \ L = 2 $}

Let us show how the general method for finding invariant solutions is applied.

\bigskip

Algebra $\mbox{\bf A}_{2,2}$
with symmetries~(\ref{Dcase23})
and invariant DODS~(\ref{DODEcase23}):

Step 1. The representative list of one-dimensional {sub}algebras~\cite{PateraWinternitz} is
$$
\{ X_1  \},
\qquad
\{ X_2  \}.
$$
Since $ X_1 $  does not satisfy condition $\xi (x,y) {\not \equiv}  0 $ we proceed only for $X_2$.

Step 2. Invariants for $X_2 $ are
\begin{equation}    \label{invariants_A22}
J_1 = { y \over x } = A ,
\qquad
J_2 = { x_-  \over x } = B .
\end{equation}

Step 3. We obtain reduction formulas
\begin{equation}        \label{solution_form_A22}
y = Ax ,
\qquad
x_- = B x  .
\end{equation}

Step 4. Substitution in~(\ref{DODEcase23}) gives
\begin{equation}        \label{restrictions_A22}
A = f (A)  ,
\qquad
B = g (A)  .
\end{equation}

Step 5. Applying group transformation $ \mbox{exp } ( \varepsilon  X_1 )$ to~(\ref{solution_form_A22}),
we obtain a more general solution
\begin{equation}     \label{general_A22}
y = Ax +  \alpha  ,
\qquad
x_- = B x  .
\end{equation}
In~(\ref{general_A22})  $\alpha $ is arbitrary, $A$ and $B$ are solutions of~(\ref{restrictions_A22}).
Thus, the number of different invariant solutions is determined
by the number of a solutions of the  equations~(\ref{restrictions_A22})
(the functions $f(z)$ and $g(z)$ in~(\ref{DODEcase23}) are given).

\bigskip

Algebra $\mbox{\bf A}_{2,4}$
with symmetries~(\ref{Dcase21})
and invariant DODS~(\ref{DODEcase21}):

The representative list   is
$$
\{   \cos ( \varphi )  X_1 +  \sin ( \varphi )   X_2 ,
\qquad
0 \leq   \varphi   < \pi \} .
$$
Invariant solutions can be obtained for $ \cos (  \varphi ) \neq 0 $.
Let    $ a =  \tan  (  \varphi )  < \infty $, then
we can rewrite the symmetry as $   X_1 +  a  X_2 $ and find invariant solutions
\begin{equation}     \label{solution_form_A24}
y  =  { a  x }  +  A   ,
\qquad
x _-  = x  - B  ,
\end{equation}
where  $a$ and $B$ satisfy
\begin{equation}        \label{restrictions_A24}
a   = f  (a)   ,
\qquad
B = g  (  B a  )
\end{equation}
and $A$  is arbitrary. Note that the invariant solutions exist only for $ a = \tan  (  \varphi ) $
which satisfies the first equation in~(\ref{restrictions_A24}).

Application  of the group transformations
$ \mbox{exp} (  \varepsilon ( \cos ( \varphi )  X_1 +  \sin ( \varphi )   X_2 ) ) $
does not give more solutions (it only changes the value of $A$).

\subsection{DODSs with $ \mbox{dim} \ L = 3 $}

For three-dimensional Lie algebras we provide all invariant solutions
for the elements
of the representative list of one one-dimensional {sub}algebras.
More general solutions which can be obtained with the help of
group transformations will be considered
only for the first two cases
$\mbox{\bf A}_{3,2}  ^a $,  $ a \neq  1 $  and $ \mbox{\bf A}_{3,4} $.

\bigskip

$\mbox{\bf A}_{3,2}  ^a $,  $ a \neq  1 $
with symmetries~(\ref{Dcase39})
and invariant DODS~(\ref{DODEcase39}):

The representative list of one-dimensional algebras is
$$
\{ X_1 \} ,
\qquad
\{ X_2 \} ,
\qquad
\{ X_1 \pm   X_2    \} ,
\qquad
\{ X_3 \}  ,
$$
where  $ X_2 $ does not satisfy the condition $\xi (x,y) { \not \equiv } 0$.
We obtain the following invariant solutions:

\begin{enumerate}

\item

For the operator $X_1$
the reduction formulas are
\begin{equation}     \label{solution_form_A32a}
y  = A   ,
\qquad
x _-  = x  - B  .
\end{equation}
The DODS~(\ref{DODEcase39}) implies $B=0$ that is not allowed.
Thus, there are no invariant solutions.

\item

For symmetry $X_1 \pm  X_2 $
the reduction formulas are
\begin{equation}     \label{solution_form_A32b}
y  =  \pm  x +   A   ,
\qquad
x _-  = x  - B   .
\end{equation}
Applying the symmetry group, we get a more general solution
\begin{equation}     \label{general_A32b}
y  =  \alpha   x +   A   ,
\qquad
x _-  = x  - B  .
\end{equation}
The DODS~(\ref{DODEcase39}) implies
\begin{equation}        \label{restrictions_A24b}
C_1  = 1  ,
\qquad
B = C_2    ( \alpha  B)   ^{1/a}     .
\end{equation}

\item

In the case of $X_3$ the solution has the form
\begin{equation}     \label{solution_form_A32c}
y  =  A  x ^a   ,
\qquad
x _-  = B x   .
\end{equation}
Applying the group transformations, we get a more general solution
\begin{equation}     \label{general_A32c}
y  - \alpha =  A ( x - \beta )  ^a   ,
\qquad
x _-  - \beta = B  (  x  - \beta )     .
\end{equation}
The conditions for constants  are
\begin{equation}        \label{restrictions_A32c}
 a   =  C_1  { 1 - B ^a \over  1 - B  }   ,
\qquad
1 - B =   C_2    \left(   A ( 1 - B^a)  \right)  ^{1/a}   .
\end{equation}

\end{enumerate}

\bigskip

$\mbox{\bf A}_{3,4}$
with symmetries~(\ref{Dcase38})
and invariant DODS~(\ref{DODEcase38}):

The representative {sub}algebras are
$$
\{ X_1 \},
\qquad
\{ X_2  \} ,
\qquad
\{ X_3 \}
$$
Note that $X_2 $ does not satisfy $\xi (x,y) {\not \equiv }  0$.

\begin{enumerate}

\item

For $ X_1 $ the invariant solution has the form
\begin{equation}     \label{solution_form_A34a}
y  = A   ,
\qquad
x _-  = x  - B  .
\end{equation}
It exists under conditions
\begin{equation}        \label{restrictions_A24a}
0   = C_1   ,
\qquad
B = C_2      .
\end{equation}

Applying the group,  we  generate more general solutions
\begin{equation}
y  = A   + \alpha x,
\qquad
x _-  = x  - B  ,
\end{equation}
which satisfy the DODE under the same conditions~(\ref{restrictions_A24a}).

\item

For symmetry $X_3$
we get the invariant solution form
\begin{equation}     \label{solution_form_A34b}
y  =  x \ln |x|   + A  x    ,
\qquad
x _-  = B x     .
\end{equation}
Group transformations extend this form to more general solutions
\begin{equation}     \label{general_A24b}
y  - \alpha  = ( x - \beta  )  \ln |x - \beta  |   + A ( x - \beta  )    ,
\qquad
x _-  - \beta  = B (x - \beta  )       ,
\end{equation}
which satisfy the DODS provided that
\begin{equation}        \label{restrictions_A24b_2}
    { B  \ln |B|  \over  1 - B  }  =  C_1   - A - 1  ,
\qquad
\mbox{sign} ( x - \beta )  ( 1 -  B )  |B| ^{ B \over 1 - B  }    =  C_2  e^A      .
\end{equation}

\end{enumerate}

\bigskip

$\mbox{\bf A}_{3,6} ^b$
with symmetries~(\ref{Dcase310})
and invariant DODS~(\ref{DODEcase310}):

The representative list of one-dimensional {sub}algebras has two elements:
$$
\{ X_1 \}   ,
\qquad
\{ X_3 \}
$$

\begin{enumerate}

\item

For $X_1$ we find invariant solutions in the form
\begin{equation}
y = A,
\qquad
x_- = x - B
\end{equation}
with conditions
$$
0 = C_1 ,
\qquad
B = C_2  .
$$

\item

For the symmetry  $X_3$ we find invariants
$$
I_1  =  e^{ b \arctan (y/x) } \sqrt{ x^2 + y^2 }   ,
\qquad
I_2  =   \arctan \left(  { y \over x }  \right)     -   \arctan \left(  { y_-  \over x_-  }  \right)   .
$$
We can use polar coordinates to represent the invariant solutions:
\begin{equation*}
x = A   e^{ - b \varphi }  \cos \varphi ,
\qquad
y = A   e^{ - b \varphi }  \sin \varphi ,
\end{equation*}
\begin{equation}
x_- = A   e^{ - b ( \varphi - B )  }  \cos ( \varphi - B )  ,
\qquad
y_-  = A   e^{ - b ( \varphi - B ) }  \sin ( \varphi - B )  .
\end{equation}
Substitution into the DODS gives us conditions on the constants:
$$
b =  {     \cos B   -   e^{ - b B }  +  C_1    \sin B    \over   \sin B   - C_1 (   \cos B   -   e^{ - b B }   ) }   ,
$$
\begin{multline*}
 \mbox{sign} ( e^{ - b \varphi }  \cos \varphi -  e^{ - b ( \varphi - B ) }  \cos  ( \varphi - B )  )  \\
\times
A \mbox{exp} \left[ b \arctan \left( {  e^{ b B }  \sin B  \over 1 - e^{ b B }  \cos B } \right)   \right]
\sqrt{  1 -  2 e^{ b B }  \cos B  + e^{ 2 b B } } = C_2   .
\end{multline*}

Let us remark that for $b=0$  the conditions are simplified to
$$
 \tan  { B \over 2}   = C_1 ,
\qquad
2    \mbox{sign} (  \cos \varphi - \cos  ( \varphi - B )  )
A \left| \sin   { B \over 2}   \right|  = C_2   .
$$

\end{enumerate}

\bigskip

$\mbox{\bf A}_{3,8}$:
For algebra $\mbox{\bf A}_{3,8}$
it is more convenient to consider the alternative version
with symmetries~(\ref{Dcase311A})
and invariant DODS~(\ref{DODEcase311A}):

The representative list of one-dimensional {sub}algebras
consists of algebras
$$
\{ X_1 \} ,
\qquad
\{ X_2 \}  ,
\qquad
\{ X_1 + X_3  \}   .
$$
We obtain the following invariant solutions:

\begin{enumerate}

\item

The {sub}algebra corresponding to the algebra $X_1 $
suggests  the invariant solution in the form
\begin{equation}
y =    A     ,
\qquad
x_- =   x  -  B
\end{equation}
provided that
$$
     { C_1 \over A }  = 0 ,
\qquad
B  = C_2   A^2   .
$$
We find only constant solutions $ y = A$,
which exist if the DODE in~(\ref{DODEcase311A})  satisfies  $C_1 = 0$.
There is one relation which gives $B$ for any value of $A$.

\item

The operator $X_2 $ implies
\begin{equation}
y =    A \sqrt{x}   ,
\qquad
x_-   =  B x
\end{equation}
and we obtain the conditions
$$
      { A  \over 2  }
      =      { A  \over 1 + \sqrt{B}  }   + { C_1  \over A }  ,
\qquad
{ 1  \over  \sqrt{B}  }   -    \sqrt{B}    = C_2  A^2
$$
for the constants.

\item

The {sub}algebra corresponding to the operator $X_1 + X_3 $
gives the invariant solution form
\begin{equation}
y =    A   \sqrt{x^2 +1}  ,
\qquad
x_- =  { x  - B \over 1 + B  x }  .
\end{equation}
Substitution into the DODS leads to the requirements
$$
{  A  \over B } ( 1 -   \mbox{sign} ( 1 + Bx)    \sqrt{ B ^2 +1}  )      +    {  C_1  \over A }   = 0  ,
\qquad
 B = C_2  A^2   \mbox{sign} ( 1 + Bx)   \sqrt{ B ^2 +1}  ,
$$
which define the constants $A$ and $B$.

\end{enumerate}

\bigskip

$\mbox{\bf A}_{3,9}$
with symmetries~(\ref{Dcase312B})
and invariant DODS~(\ref{DODEcase312B}):

The representative list has the same form as in the previous case:
$$
\{ X_1 \} ,
\qquad
\{ X_2 \} ,
\qquad
\{ X_1 + X_3  \}
$$
We consider the operators satisfying $ \xi (x,y) {\not \equiv} 0$.

\begin{enumerate}

\item

For operator  $X_2$ we get invariant solutions
\begin{equation}
y = A x ,
\qquad
x_- = B x
\end{equation}
under the conditions
$$
A = {     A^2  ( 1 - B )^2  +  B^2  -  1  + 2 C_1 A ( 1 - B )   \over 2 A ( 1 - B ) - C_1 ( A^2 ( 1 - B ) ^2 +  B^2 - 1 )   }  ,
$$
$$
( A^2 +  1 ) ( 1 -  B )  ^2 = C_2 B   .
$$

\item

For operator $X_1 + X_3 $ we get invariants
$$
I_1 = { y^2 + x^2 +1 \over x } ,
\qquad
I_2 = \arctan \left(   { y^2 + x^2 - 1 \over 2 y  }  \right)
-   \arctan \left(   { y_- ^2 + x_- ^2 - 1 \over 2 y_-  }      \right)   .
$$
In view of the complicated form of the invariants we will not
present the solutions here and in any case the equations cannot be solved explicitly.

\end{enumerate}

\bigskip

$\mbox{\bf A}_{3,10}$:
In the case $\mbox{\bf A}_{3,10}$ we consider the alternative realization
with symmetries~(\ref{Dcase313C})
and invariant DODS~(\ref{DODEcase313C})

The representative list is
$$
\{ X_1 \}  ,
\qquad
\{ X_2 \}  ,
\qquad
\{ X_1 + X_3  \}
$$
as for the previous two realizations of  $  {\sso {sl}}   (2, \mathbb{F})$.
We obtain invariant solutions for all three elements.

\begin{enumerate}

\item

Invariant solutions for $ X_1 $ have the form
\begin{equation}
y = x + A   ,
\qquad
x_-  =  x  - B
\end{equation}
provided that
$$
1  = C_1 \left( { A +  B   \over   B  } \right) ^2  ,
\qquad
{  B  ^2   \over  A  ^2   }  = C_2   .
$$

\item

For $ X_2 $ we look for invariant solution in the form
\begin{equation}
y = A x  ,
\qquad
x_-  = B x
\end{equation}
and obtain the system
$$
A  = C_1 \left( { A - B   \over   1 - B  } \right) ^2  ,
\qquad
{  A   ( 1 - B ) ^2   \over  (A - 1 ) ^2  B  } = C_2  ,
$$
which defines the constants $A$ and $B$.

\item

For $ X_1 + X_3 $ we look for invariant solutions in the form
\begin{equation}
y = { x + A  \over  1 - A x } ,
\qquad
x_-  = { x - B  \over  1 + B x }
\end{equation}
and get restrictions
$$
1 + A^2   = C_1 \left( 1 + {  A \over B }  \right) ^2 ,
\qquad
{  ( 1 + A^2 ) B^2   \over   A^2 ( 1 + B^2 )  } = C_2   .
$$

\end{enumerate}

\bigskip

$\mbox{\bf A}_{3,12}$
with symmetries~(\ref{Dcase314})
and invariant DODS~(\ref{DODEcase314}):

The representative list contains only one element
$$
\{ X_1  \}   .
$$
Invariant solutions must have the form
\begin{equation}
y = A \sqrt{ 1 + x^2 } ,
\qquad
x_- = {  x - B  \over 1 +  B x }
\end{equation}
and get the restrictions
$$
{   A      \over  { 1 + A^2 }    }
\left( 1 -  {  \mbox{sign} ( 1 +  B x )  \over    \sqrt{ 1 + B^2 } } \right)  = C_1   ,
$$
$$
\left(  {  \mbox{sign} ( 1 +  B x )  \over    \sqrt{ 1 + B^2 } }   + A^2   \right)  ^2
= ( 1 - C_2 )   ( 1 + A^2 ) ^2   .
$$

\section{Conclusions}     \label{Conclusions}

In this paper we presented a Lie group classification of delay ordinary differential
systems, which consist of first-order DODEs and delay relations. This approach
has some common features with the Lie group classification of ordinary difference
equations supplemented by lattice equations (ordinary difference systems).
However solutions of delay differential systems are continuous while solutions
of ordinary difference systems exist only in lattice points.

The dimension of a symmetry algebra admitted by a DODS can be
$n = 0, \ 1, \ 2, \ 3$, or $n = \infty$.
If $n = \infty$,  the DODS consists of a linear DODE and a
solution independent delay relation, or it can be transformed into such a form by an
invertible transformation. Genuinely nonlinear DODS can admit symmetry algebras
of dimension $0 \leq  n \leq  3$. The results are summed up in the Table 2.
Analyzing the classification, we obtained several theoretical results.
Namely, if the symmetry algebra has two linearly connected symmetries, it provides a delay
differential system which can be transformed into a linear DODE supplemented
by a solution independent delay relation (see Theorem~\ref{lineartheorem}).
Such linear delay
differential systems admit infinite-dimensional symmetry groups since they allow
linear superposition of solutions (Theorem~\ref{th41a}).

In this respect DODSs are similar to PDEs rather than ODEs.
The reason is that initial conditions for ODEs are given in one point
and hence consist of a finite number of constants.
For DODSs (and PDEs) initial conditions must be given on an interval
and hence involve arbitrary functions
(in our case one function of one variable).

   Linear DODSs~(\ref{linearDODE}) split into two classes.
In one of them there are DODSs which in
addition to  the infinite-dimensional symmetry algebra
corresponding to the superposition principle admit one further element~(\ref{additional}).
The “superposition group” acts only on the dependent variable $y$
whereas the additional symmetry also acts on the independent variable $x$.
Such DODSs can be brought into the form~(\ref{canonical1new}). 
If we know one particular solution, we can transform the DODE into the homogeneous form 
and bring the DODS to the form~(\ref{canonical1}).
This class of linear DODSs is characterized by the fact that
the coefficients $\alpha(x) $, $\beta(x)$ and $g(x)$ satisfy the constraint~(\ref{compatibil}).

 Systems in the other class do not admit any additional symmetry
not related to the linear superposition symmetry.
They can be transformed into the form~(\ref{canonical2new}), or equivalently~(\ref{canonical3new}).
If we know one particular solution, we can transform the DODE into the homogeneous form 
and bring the DODS to the form~(\ref{canonical2}), or equivalently~(\ref{canonical3}).

   In Section~\ref{invariantsolutions1},
as an application of the symmetries admitted by DODS
we presented a procedure for calculating particular solutions,
which are invariant solutions with respect to one-dimensional {sub}groups of the symmetry group.
We applied it to all nonlinear representative DODSs found in our classification in Section~\ref{Classification}.
In order to provide invariant solutions the symmetry algebra must contain
at least one element with $\xi  (x,y) {\not \equiv } 0  $  in~(\ref{operator1}).

  Linear DODSs will be treated in detail in a separate article.

\bigskip
\medskip

\noindent
{ \bf Acknowledgments}

 The research of VD was partly supported by research grant
No. 15-01-04677-A of the Russian Fund for Base Research.
The research of PW was partially supported by a research grant from NSERC of Canada.

\eject

\noindent
{\bf {\large Appendix.}}

\bigskip

\begin{center}
{\bf  Table~1. Lie symmetry algebras and their realizations}
\end{center}

In Column~1 we give the isomorphism class using the notations of~\cite{SWbook}.
Thus $ {\sso {n}}_{i,k} $ denotes the $k$-th nilpotent Lie algebra of dimension $i$ in the list.
The only nilpotent algebras in Table~1 are $ {\sso {n}}_{1,1} $, $ {\sso {n}}_{3,1} $ and $ {\sso {n}}_{4,1} $.
Similarly, $ {\sso {s}}_{i,k} $ is the $k$-th solvable Lie algebra of dimension $i$ in the list.
In Column~2 $ \mbox{\bf A}_{i,k} $ runs through all algebras in the list of sub{algebras}
of $ \mbox{diff} ( 2, \mathbb{R}  ) $ and $i$ is again the dimension of the algebra.
Simple Lie algebras are identified by their usual names
($ {\sso {sl}} ( 2, \mathbb{R}  ) $, $ {\sso {o}} ( 3 ) $).
The numbers in brackets correspond to notations used in the list of Ref.~\cite{Gonzalez1992}.
In Column~3 we give vector fields spanning each representative algebra.

\bigskip


\noindent
{\bf Dimensions 1 and 2}

$$
\begin{array}{|c|c|l|}
\hline
  &  &    \\
\mbox{Lie algebra} & \mbox{Case} & \mbox{Operators} \\
  &  &    \\
\hline
  &  &    \\
{\sso {n}} _{1,1}
&
\mbox{\bf A}_{1,1} (9)
&
{ \displaystyle   X_1 = { \ddy}   }
\\
  &  &    \\
\hline
    &  &   \\
{\sso {s}} _{2,1}
&
\mbox{\bf A}_{2,1} (10)
&
{ \displaystyle
X_1 = { \ddy}   ;
\quad
X_2 =  y  { \ddy} }
\\
    &   &   \\
\cline{2-3}
    &   &  \\
&
\mbox{\bf A}_{2,2} (22)
&
{ \displaystyle
X_1 = { \ddy}   ;
\quad
X_2 = x { \ddx}  +   y  { \ddy}  }
\\
  &     &  \\
\hline
  &  &    \\
2 {\sso {n}} _{1,1}
&
\mbox{\bf A}_{2,3} (20)
&
{ \displaystyle
\left\{  X_1 = { \ddy}  \right\} ,
\quad
\left\{  X_2 = x  { \ddy}  \right\}  }
\\
    &      &   \\
\cline{2-3}
    &  &  \\
&
\mbox{\bf A}_{2,4} (22)
&
{ \displaystyle
\left\{  X_1 = { \ddx} \right\} ,
\quad
\left\{  X_2 = { \ddy} \right\}  }
\\
   &      &  \\
\hline
\end{array}
$$

\eject

\noindent
{\bf Dimension 3}

$$
\begin{array}{|c|c|l|}
\hline
  &   &    \\
\mbox{Lie algebra} & \mbox{Case} & \mbox{Operators}  \\
  &   &    \\
\hline
  &   &    \\
{\sso {n}} _{3,1}
&
\mbox{\bf A}_{3,1} (22)
&
 { \displaystyle
X_1 = { \ddy} ;
\quad
X_2 =  x  { \ddy} ,
\quad
X_3 = { \ddx} }
\\
  &     &    \\
\hline
  &  &    \\
{\sso {s}} _{3,1}
&
\mbox{\bf A}_{3,2} ^a  (12)
&
{ \displaystyle X_1 =  { \ddx} ,
\quad
X_2 =  { \ddy}  ;
\quad
X_3 =   x  { \ddx}  +  a  y  { \ddy}  ,
\quad
0 < |a| \leq 1
  }
\\
  &    &    \\
\cline{2-3}
  &  &    \\
&
\mbox{\bf A}_{3,3} ^a  (21, 22)
&
{ \displaystyle
X_1 = { \ddy} ,
\quad
X_2 = x { \ddy} ;
\quad
X_3 =  (1-a) x { \ddx}  +  y { \ddy} ,
\quad
0 < |a| \leq 1
  }
\\
  &    &   \\
\hline
   &  &   \\
{\sso {s}} _{3,2}
&
\mbox{\bf A}_{3,4}   (25)
&
{ \displaystyle
X_1 = { \ddx} ,
\quad
X_2 =  { \ddy} ;
\quad
X_3 =   x  { \ddx}  +   ( x  +  y )  { \ddy}   }
\\
   &     &    \\
\cline{2-3}
   &  &    \\
&
\mbox{\bf A}_{3,5}  (22)
&
 { \displaystyle
X_1 = { \ddy} ,
\quad
X_2 = x { \ddy} ;
\quad
X_3 =  { \ddx}  +  y { \ddy}   }
\\
   &     &   \\
\hline
   &  &   \\
{\sso {s}} _{3,3}
&
\mbox{\bf A}_{3,6}  ^b  (1)
&
 {\displaystyle
X_1 =  { \ddx} ,
\quad
X_2 =  { \ddy} ;
\quad
X_3 =   (b x + y )  { \ddx}  +  ( b  y - x )   { \ddy}  ,
\quad
b \geq 0 }
\\
   &   &   \\
\cline{2-3}
   &  &   \\
&
\mbox{\bf A}_{3,7}  ^b   (22)
&
{ \displaystyle
X_1 = { \ddy} ,
\quad
X_2 = x { \ddy} ;
\quad
X_3 =   ( 1   +  x ^2 )  { \ddx}   +  ( x + b)   y { \ddy}  ,
\quad
b \geq 0  }
\\
  &     &   \\
\hline
\end{array}
$$

$$
\begin{array}{|c|c|l|}
\hline
  &  &     \\
\mbox{Lie algebra} & \mbox{Case} & \mbox{Operators}   \\
  &  &    \\
\hline
  &  &    \\
{\sso {sl}}  (2, \mathbb{R})
&
\mbox{\bf A}_{3,8}  (18)
&
  {\displaystyle
X_1 = { \ddy} ,
\quad
X_2 =  x { \ddx} +  y   { \ddy} ,
\quad
X_3 =   2 x y  { \ddx}  +   y ^2  { \ddy}     }
\\
  &     &    \\
\cline{2-3}
  &   &    \\
&
\mbox{\bf A}_{3,9}  (2)
&
   {\displaystyle
X_1 = { \ddy}  ,
\quad
X_2 =  x  { \ddx}  +  y  { \ddy}  ,
\quad
X_3 =   2 x y  { \ddx}  +   ( y ^2  - x^2 )  { \ddy}  }
\\
  &     &     \\
\cline{2-3}
  &   &    \\
&
\mbox{\bf A}_{3,10}  (17)
&
 {\displaystyle
X_1 = { \ddy} ,
\quad
X_2 =  x  { \ddx}  +  y  { \ddy} ,
\quad
X_3 =   2 x y  { \ddx}  +   ( y ^2  + x^2 )   { \ddy}    }
\\
  &    &    \\
\cline{2-3}
  &  &   \\
&
\mbox{\bf A}_{3,11}  (11)
&
 {\displaystyle
X_1 =   { \ddy} ,
\quad
X_2 = y  { \ddy} ,
\quad
X_3 = y^2  { \ddy}   }
\\
  &      &   \\
\hline
  &  &    \\
{\sso {o}}  (3, \mathbb{R})
&
\mbox{\bf A}_{3,12}   (3)
&
 {\displaystyle
X_1 = ( 1 + x^2  )  { \ddx} + xy  { \ddy} ,
\quad
X_2 = xy  { \ddx} + ( 1 + y^2  )  { \ddy} ,
}
\\
  &  &   \\
&
&
 {\displaystyle
X_3 = y  { \ddx}  -  x  { \ddy}   }
\\
  &   &   \\
\hline
  &  &    \\
{\sso {n}}_{1,1}    \oplus {\sso {s}}_{2,1}
&
\mbox{\bf A}_{3,13}  (23)
&
{ \displaystyle
\left\{ X_1 = { \ddx} \right\} ,
\quad
\left\{
X_2 = { \ddy} ;
\quad
X_3 =  y { \ddy}
\right\}  }
\\
  &  &    \\
\cline{2-3}
 &  &    \\
&
\mbox{\bf A}_{3,14}   (22)
&
{ \displaystyle
\left\{  X_1 = x { \ddy}  \right\}   ,
\quad
\left\{
X_2 = { \ddy} ;
\quad
X_3 =   x { \ddx}  + y { \ddy}
\right\}   }
\\
  &   &    \\
\hline
  &  &    \\
3{\sso {n}}_{1,1}
&
\mbox{\bf A}_{3,15}  (20)
&
{\displaystyle
\left\{  X_1 =   { \ddy} \right\}  ,
\quad
\left\{  X_2 = x   { \ddy} \right\} ,
\quad
\left\{  X_3 =  \chi (x)   { \ddy}  \right\}  ,
\quad
\ddot{\chi}   (x) {\not\equiv}  0  }
\\
  &    &     \\
\hline
\end{array}
$$

\eject

\noindent
{\bf Dimension 4}



$$
\begin{array}{|c|c|l|}
\hline
  &   &     \\
\mbox{Lie algebra} & \mbox{Case} & \mbox{Operators}   \\
  &   &     \\
\hline
 &  &   \\
{\sso {n}}_{4,1}
&
\mbox{\bf A}_{4,1} (22)
&
  {\displaystyle
X_1 =   { \partial \over \partial y  } ,
\quad
X_2 =  x  { \partial \over \partial y  } ;
\quad
X_3 =  x ^2    { \partial \over \partial y  }   ,
\quad
X_4 = { \partial \over \partial x }
 }
\\
  &    &   \\
\hline
  &  &   \\
{\sso {s}}_{4,1}
&
\mbox{\bf A}_{4,2}  (22)
&
  {\displaystyle
X_1 =    { \partial \over \partial y  } ,
\quad
X_2 =  x    { \partial \over \partial y  } ,
\quad
X_3 =  e ^{x}   { \partial \over \partial y  } ;
\quad
X_4 = { \partial \over \partial x }
}
\\
  &    &   \\
\hline
 &  &   \\
{\sso {s}}_{4,2}
&
\mbox{\bf A}_{4,3}  (22)
&
  {\displaystyle
X_1 =   { \partial \over \partial y  } ,
\quad
X_2 =  x  { \partial \over \partial y  } ,
\quad
X_3 =  x ^2    { \partial \over \partial y  }  ;
\quad
X_4 = { \partial \over \partial x }  + y { \partial \over \partial y  }
 }
\\
  &   &   \\
\hline
 &  &   \\
{\sso {s}}_{4,3}
&
\mbox{\bf A}_{4,4}  ^{a, \alpha}   (22)
&
  {\displaystyle
X_1 =    { \partial \over \partial y  } , \quad X_2 =  x { \partial
\over \partial y  } , \quad X_3 =   |x|  ^{ \alpha }   { \partial
\over \partial y  }  ; \quad X_4 = ( 1 - a) x { \partial \over
\partial x }  + y   { \partial \over \partial y } ,
}
\\
  &  &   \\
&  &
{\displaystyle
a \in [-1 , 0) \cup  (0,1)  ,
\
 \alpha  \neq  \{  0, {1 \over 1-a} , 1 \}   }
\\
  &  &   \\
&  &
{\displaystyle
\mbox{(see~\cite{SWbook} for additional restriction on  $a$ and $\alpha$)} }
\\
  &   &   \\
\cline{2-3}
  &  &   \\
&
\mbox{\bf A}_{4,5} (21)
&
  {\displaystyle
X_1 = { \partial \over \partial y }   ,
\quad
X_2 =  x  { \partial \over \partial y  } ,
\quad
X_3 =  \chi (x) { \partial \over \partial y  } ;
\quad
X_4 = y { \partial \over \partial y }   ,
\quad
\ddot{\chi}(x) {\not\equiv} 0   }
\\
  &  &   \\
\hline
  &  &   \\
{\sso {s}}_{4,4}
&
\mbox{\bf A}_{4,6} ^a (22)
&
  {\displaystyle
X_1 =    { \partial \over \partial y  } ,
\quad
X_2 =  x    { \partial \over \partial y  } ,
\quad
X_3 =  e ^{ a x}   { \partial \over \partial y  } ;
\quad
X_4 = { \partial \over \partial x }  +  y { \partial \over \partial y  } ,
\quad
a \neq 0 , 1 }
\\
  &  &   \\
\hline
 &  &   \\
{\sso {s}}_{4,5}
&
\mbox{\bf A}_{4,7}  ^ {\alpha, \beta}  (22)
&
  {\displaystyle
X_1 =  { \partial \over \partial y  } ,
X_2 =  e ^{ \alpha  x} \cos ( \beta x )  { \partial \over \partial y  } ,
X_3 =  e ^{ \alpha x} \sin ( \beta x )  { \partial \over \partial y  } ;
X_4 = { \partial \over \partial x }   + y  { \partial \over \partial y  }  }, \\
  &  &   \\
 &  &
 \beta \neq 0
  \\
  &   &   \\
\hline
\end{array}
$$

\eject

$$
\begin{array}{|c|c|l|}
\hline
  &   &     \\
\mbox{Lie algebra} & \mbox{Case} & \mbox{Operators}   \\
  &   &     \\
\hline
   &  &   \\
{\sso {s}}_{4,6}
&
\mbox{\bf A}_{4,8}  (24)
&
  {\displaystyle
X_1 =  { \partial \over \partial y }   ,
\quad
X_2 = { \partial \over \partial x }   ,
\quad
X_3 = x  { \partial \over \partial y  }  ;
\quad
X_4 =  x  { \partial \over \partial x  }
  }
\\
  &  &    \\
\hline
  &  &   \\
{\sso {s}}_{4,8}
&
\mbox{\bf A}_{4,9} ^a  (24)
&
  {\displaystyle
X_1 =  { \partial \over \partial y }   ,
\quad
X_2 = { \partial \over \partial x }   ,
\quad
X_3 = x  { \partial \over \partial y  }  ;
\quad
X_4 =  x  { \partial \over \partial x  }   +  a y  { \partial \over \partial y  } ,
\quad
a \neq  0,1
  }
\\
  &  &    \\
\hline

  &  &    \\
{\sso {s}}_{4,10}
&
\mbox{\bf A}_{4,10}  (25)
&
  {\displaystyle
X_1 =  { \partial \over \partial y }   ,
\quad
X_2 = { \partial \over \partial x }   ,
\quad
X_3 = x  { \partial \over \partial y  }  ;
\quad
X_4 =    x { \partial \over \partial x } +    ( 2 y + x^2 )  { \partial \over \partial y  } }
\\
  &   &    \\
\hline
   &  &   \\
{\sso {s}}_{4,11}
&
\mbox{\bf A}_{4,11} (24)
&
  {\displaystyle
X_1 =  { \partial \over \partial y }   ,
\quad
X_2 = { \partial \over \partial x }   ,
\quad
X_3 = x  { \partial \over \partial y  }  ;
\quad
X_4 =  x  { \partial \over \partial x  }   +  y  { \partial \over \partial y  }
  }
\\
  &  &    \\
\cline{2-3}
  &  &   \\
&
\mbox{\bf A}_{4,12} (23)
&
  {\displaystyle
X_1 =  { \partial \over \partial y }   ,
\quad
X_2 =  x  { \partial \over \partial y  } ,
\quad
X_3 =  { \partial \over \partial x  }  ;
\quad
X_4 =  y { \partial \over \partial y  }
 }
\\
  &  &   \\
\hline
 &   &     \\
{\sso {s}}_{4,12}
&
\mbox{\bf A}_{4,13}  (4)
&
  {\displaystyle
X_1 = { \partial \over \partial x }  ,
\quad
X_2 =   { \partial \over \partial y } ;
\quad
X_3 = x  { \partial \over \partial x }  + y { \partial \over \partial y } ,
\quad
X_4 = y  { \partial \over \partial x }  -  x  { \partial \over \partial y } }
\\
 &   &   \\
\cline{2-3}
 &  &   \\
&
\mbox{\bf A}_{4,14}   (23)
&
  {\displaystyle
X_1 =  { \partial \over \partial y  } ,
\quad
X_2 =  x  { \partial \over \partial y  } ;
\quad
X_3 =  y { \partial \over \partial y  } ,
\quad
X_4 = ( 1 + x^2 ) { \partial \over \partial x } +  x y { \partial \over \partial y }
  }
\\
  &   &   \\
\hline
\end{array}
$$

\eject

$$
\begin{array}{|c|c|l|}
\hline
  &   &     \\
\mbox{Lie algebra} & \mbox{Case} & \mbox{Operators}   \\
  &   &     \\
\hline
  &  &   \\
{\sso {n}}_{1,1} \oplus {\sso {s}}_{3,1}
&
\mbox{\bf A}_{4,15} ^a  (22)
&
  {\displaystyle
\left\{X_1 =   |x|  ^{1 \over 1-a}   { \partial \over \partial y  }
\right\},
\left\{
X_2 =    { \partial \over \partial y  } ,
X_3 =  x { \partial \over \partial y  } ;
X_4 = ( 1 - a) x { \partial \over \partial x }  + y   { \partial \over \partial y }
\right\} ,
}
\\
  &  &   \\
&  &
{\displaystyle
a \in [-1 , 0) \cup  (0,1)   }
\\
  &   &   \\
\hline
  &  &   \\
{\sso {n}}_{1,1} \oplus {\sso {s}}_{3,2}
&
\mbox{\bf A}_{4,16} (22)
&
  {\displaystyle
\left\{ X_1 =  e ^{x}   { \partial \over \partial y  } \right\},
\quad
\left\{
X_2 =    { \partial \over \partial y  } ,
\quad
X_3 =  x    { \partial \over \partial y  } ;
\quad
X_4 = { \partial \over \partial x }  + y { \partial \over \partial y }
\right\}
  }
\\
  &  &   \\
\hline
  &  &   \\
{\sso {n}}_{1,1} \oplus {\sso {s}}_{3,3}
&
\mbox{\bf A}_{4,17} ^{ \alpha } (22)
&
  {\displaystyle
\left\{  X_1 = { \partial \over \partial y }   \right\}   ,
\left\{
X_2 =   { \partial \over \partial x  } ,
X_3 =  e ^{ \alpha  x} \cos  x   { \partial \over \partial y  } ,
X_4 =  e ^{ \alpha x}  \sin   x   { \partial \over \partial y  }
\right\} ,
} \\
  &   &   \\
  &   &  { \alpha  \geq 0 }
\\
  &   &   \\
\hline
 &  &   \\
{\sso {n}}_{1,1} \oplus {\sso {sl}}  (2, \mathbb{R})
&
\mbox{\bf A}_{4,18} (14)
&
  {\displaystyle
\left\{  X_1 = { \partial \over \partial x } \right\}   ,
\quad
\left\{
X_2 =  { \partial \over \partial y } ,
\quad
X_3 = y { \partial \over \partial y }  ,
\quad
X_4  = y^2 { \partial \over \partial y }
\right\}  }
\\
 &     &   \\
\cline{2-3}
 &   &   \\
&
\mbox{\bf A}_{4,19}  (19)
&
  {\displaystyle
\left\{ X_1 =   x { \partial \over \partial x } \right\} ,
\left\{
X_2 = { \partial \over \partial y }   ,
X_3 = x { \partial \over \partial x }  +   y { \partial \over \partial y } ,
X_4 = 2 x y   { \partial \over \partial x }  +  y^2   { \partial \over \partial y }
\right\}  }
\\
  &    &   \\
\hline
  &   &  \\
2 {\sso {s}}_{2,1}            
&
\mbox{\bf A}_{4,20} (13)
&
  {\displaystyle
\left\{
X_1 = { \partial \over \partial x } ;
\quad
X_2 = x { \partial \over \partial x }
\right\} ,
\quad
\left\{
X_3 =  { \partial \over \partial y }  ;
\quad
X_4  = y { \partial \over \partial y }
\right\} }
\\
  &   &   \\
\cline{2-3}
  &  &   \\
&
\mbox{\bf A}_{4,21}  (23)
&
  {\displaystyle
\left\{
X_1 =   { \partial \over \partial y  } ;
\quad
X_2 = x { \partial \over \partial x }  +   y { \partial \over \partial y  }
\right\} ,
\quad
\left\{
X_3 =  x { \partial \over \partial y  } ;
\quad
X_4 =  x { \partial \over \partial x  }
\right\} }
\\
  &   &   \\
\hline
 &  &   \\
4 {\sso {n}}_{1,1}
&
\mbox{\bf A}_{4,22}  (20)
&
  {\displaystyle
\left\{  X_1 = { \partial \over \partial y }  \right\}  ,
\left\{  X_2 = x { \partial \over \partial y  } \right\} ,
\left\{  X_3 =  \chi_1 (x) { \partial \over \partial y  } \right\} ,
\left\{  X_4 =  \chi_2 (x) { \partial \over \partial y  } \right\}  }
\\
  &  &   \\
&
&
\mbox{$1$, $x$, $\chi_1 (x)$ and $\chi_2 (x) $ are linearly independent}
\\
  &   &   \\
\hline
\end{array}
$$

\eject

\begin{center}
{\bf Table~2. Classification of nonlinear invariant DODSs}
\end{center}



$$
\begin{array}{|c|c|c|}
\hline
    &   &  \\
\mbox{Case} & \mbox{DODE} & \mbox{Delay relation}  \\
    &   &  \\
\hline
    &   &  \\
\mbox{\bf A}_{1,1}
&
{ \displaystyle
\dot{y} = f \left( x, { \Delta y \over  \Delta x }  \right)
}
&
 \Delta x  =   g ( x, \Delta y  )
\\
    &   &  \\
\hline
   &   &  \\
\mbox{\bf A}_{2,2}
&
{ \displaystyle  \dot{y}   =    f  \left( {  \Delta y    \over     \Delta   x   }  \right) }
&
 { \displaystyle   x_-    =    x   g  \left( {  \Delta y    \over     \Delta   x   }  \right)   }
\\
    &   &  \\
\hline
    &   &  \\
\mbox{\bf A}_{2,4}
&
 { \displaystyle  \dot{y}  =  f  \left(  {  \Delta y    \over  \Delta x  }  \right)   }
&
 \Delta   x = g (  \Delta y  )
\\
    &   &  \\
\hline
\end{array}
$$

$$
\begin{array}{|c|c|c|}
\hline
    &   &  \\
\mbox{Case} & \mbox{DODE} & \mbox{Delay relation}  \\
    &   &  \\
\hline
    &   &  \\
\mbox{\bf A}_{3,2}  ^a  \quad a \neq 1
&
{ \displaystyle   \dot{y}  =   C_1 { \Delta y  \over \Delta x  }      }
&
 \Delta  x      =  
 (  \Delta y  ) ^{1/a}
\\
    &   &  \\
\hline
    &   &  \\
\mbox{\bf A}_{3,2}  ^a  \quad a = 1
&
\mbox{No DODE}
&
\mbox{or no delay relation }    \\
    &   &  \\
\hline
    &   &  \\
\mbox{\bf A}_{3,4}
&
{ \displaystyle   \dot{y}  =   { \Delta y  \over \Delta x  }    +   C_1  }
&
 { \displaystyle  \Delta  x    =  C_2   e^ { \Delta y  \over \Delta x  } }
\\
    &   &  \\
\hline
    &   &  \\
\mbox{\bf A}_{3,6} ^b
&
  \dot{y}
= {  {\displaystyle  {  \Delta y  \over  \Delta x   }      + C_1  }
 \over
 {\displaystyle  1 - C_1 {  \Delta y  \over  \Delta x   }  }  }
&
 {\displaystyle   \Delta x     e^{ b   \arctan    {  \Delta y  \over  \Delta x   }   }
\sqrt{  1 + \left( {  \Delta y  \over  \Delta x   }  \right) ^2 }  =  C_2   }
\\
    &   &  \\
\hline
    &   &  \\
\mbox{\bf A}_{3,8}
&
  {\displaystyle   \dot{y}     =  {  \Delta y   \over 2 x + C_1  \Delta y   }      }
&
    x x_-     =  C_2   (\Delta y)^2
\\
    &   &  \\
\hline
    &   &  \\
\mbox{\bf A}_{3,9}
&
   {\displaystyle     \dot{y}
= {       { ( \Delta y )^2  + x_- ^2  - x ^2 } + 2 C_1 x \Delta y
\over
2x \Delta y   - C_1  (  { ( \Delta y )^2  + x_- ^2  - x ^2 }  )  }    }
&
{ ( \Delta y )^2  + ( x  - x_- )^2  }       =  C_2    x x_-
\\
    &   &  \\
\hline
    &   &  \\
\mbox{\bf A}_{3,10}
&
 {\displaystyle     \dot{y}
= {  {  ( \Delta y  )^2  + x^2 - x_- ^2  }      + 2 C_1 x  \Delta y
\over
 2 x  \Delta y  + C_1 ( {  ( \Delta y  )^2  + x^2 - x_- ^2  }   ) }    }
&
{ ( \Delta y )^2  - ( x  - x_- )^2   }       =  C_2     x x_-
\\
    &   &  \\
\hline
    &   &  \\
\mbox{\bf A}_{3,12}
&
 {\displaystyle     {
 ( x - x _- )
\left( \dot{y} -   {\displaystyle  { y - y _- \over x - x_- } } \right)
\over
\sqrt{  1 +  \dot{y} ^2 +  ( y - x  \dot{y} ) ^2  }
\sqrt{  1 + x _- ^2 +    y _- ^2  }
}
= C_1   }
&
{
 ( x - x _- )^2
\left( 1  + \left(  { y - y _- \over x - x_- }   \right)^2
+ \left( y - x {   y - y _- \over  x - x_- } \right)^2   \right)
\over
(  1 + x^2 +    y ^2  )
(  1 + x _- ^2 +    y _- ^2  )
}
=  C_2
\\
    &   &  \\
\hline
\end{array}
$$

 \end{document}